\documentclass[11pt]{article}

\usepackage{amsmath}
\usepackage{amssymb}
\usepackage{fullpage}
\usepackage{graphicx}
\usepackage{bm}% bold math
\usepackage{hyperref}% add hypertext capabilities
\usepackage{cite}
\usepackage{authblk}

\PassOptionsToPackage{hyphens}{url}\usepackage{hyperref}

\title{\textbf{Evolution of cooperation and consistent personalities in public goods games \footnote{Published in Scientific Reports, Sci Rep 11, 23708 (2021). \href{https://doi.org/10.1038/s41598-021-03045-w}{https://doi.org/10.1038/s41598-021-03045-w}}}}
\date{}

\author{\textbf{Mohammad Salahshour}\thanks{\texttt{salahshour.mohammad@gmail.com}.}}
\affil{Max Planck Institute for Mathematics in the Sciences, Inselstrasse 22, D-04103, Leipzig, Germany}

\begin{document}
\maketitle

\begin{abstract}
The evolution of cooperation has remained an important problem in evolutionary theory and social sciences. In this regard, a curious question is why consistent cooperative and defective personalities exist and if they serve a role in the evolution of cooperation? To shed light on these questions, here, I consider a population of individuals who possibly play two consecutive rounds of public goods game, with different enhancement factors. Importantly, individuals have independent strategies in the two rounds. However, their strategy in the first round affects the game they play in the second round. I consider two different scenarios where either only first-round cooperators play a second public goods game, or both first-round cooperators and first-round defectors play a second public goods game, but in different groups. The first scenario can be considered a reward dilemma, and the second can be considered an assortative public goods game but with independent strategies of the individuals in the two rounds. Both models show cooperators can survive either in a fixed point or through different periodic orbits. Interestingly, due to the emergence of a correlation between the strategies of the individuals in the two rounds, individuals develop consistent personalities during the evolution. This, in turn, helps cooperation to flourish. These findings shed new light on the evolution of cooperation and show how consistent cooperative and defective personalities can evolve and play a positive role in solving social dilemmas.
\end{abstract}
\section*{Introduction}

As cooperation is costly, an individual is better off not cooperating. This leads to a tragedy of the commons where everyone ends up being worse off than if otherwise, all had cooperated \cite{Hardin}. However, contrary to what a rational argument suggests, empirical evidence shows that tragedies of the commons are not that common in nature \cite{Boyd}. Over the past decades, many efforts have been devoted to understanding how evolution has prevented tragedies of the commons and promoted high levels of cooperation \cite{Boyd,Axelrod,Nowak,Perc2013}. Prisoner's dilemma, and its extension to $n$ players, public goods game (PGG), have been the most common frameworks in many of these studies \cite{Doebeli,Axelrod,Nowak,Perc2013,Axelrod}. In the latter game, each player in a group of $n$ players can decide whether to invest an amount $c$ in a public good or not. All the investments are multiplied by an enhancement factor $r<n$ and are divided equally among the players. Defectors, refraining from investment, receive the highest payoff and are expected to dominate the population. This expectation contradicts observation \cite{Boyd,Axelrod}. Past researches have revealed different mechanisms through which this puzzle can be solved. For instance, when interactions are repeated, cooperation can evolve due to the threat of retaliation of a defective act in future rounds \cite{Hilbe2018a,Axelrod,Schmid2021}. However, in many cases, interactions are not repeated. A similar mechanism, indirect reciprocity \cite{Milinski,Panchanathan,Schmid2021,Alexander2017}, based on which an individual's reputation determines others' behavior towards the individual can be at work to promote cooperation in non-repeated interactions. Similarly, punishment of defectors \cite{Fehr,Perc,Boyd2,Szolnoki,Hauert1,Hilbe2,Salahshour2,Salahshour2021B} or rewarding cooperators \cite{Rand,Attila,Hilbe} can promote social behavior. Cooperation can also evolve when interactions are not obligatory \cite{Hauert,Szabo}, or when individuals have a choice between different institutions \cite{Salahshour0,Salahshour2021A}. Furthermore, it is shown that the very existence of population structure can promote cooperation due to the assortativity of interactions in structured populations \cite{Ohtsuki2006,Perc2013}. Other studies have shown assortativity \cite{Iyer2020}, for instance resulting from kin selection, group selection \cite{Nowak} or tag-based mechanisms \cite{Riolo2001}, social diversity \cite{Qin,Santos}, heterogeneity \cite{Perc2011,Stilwell2020,Kun2013}, conformity \cite{Hu2019,Szolnoki2015}, costly signaling \cite{Salahshour1,Gintis}, moral norms \cite{Salahshour2021C,Alexander2017}, and coevolution of cooperation and language \cite{Salahshour2020}, to mention a few, can play a positive role in the evolution of cooperation.

Despite the valuable insights reached on the subject, open questions regarding the mechanisms and conditions under which cooperative behavior is expected to flourish, remain to be addressed. A curious question in this regard is the existence of consistent cooperative and defective personalities. For instance, public goods experiments have shown that while about half of the people are conditional cooperators who are willing to cooperate provided their group-mates cooperate, others tend to free-ride on others' contributions \cite{Fischbacher1,Fischbacher2,Burlando,Chaudhuri}. Similar observations regarding consistent personality differences in humans and animals, in different contexts, have been made \cite{Bergmuller}. These observations raise the question of how such consistent cooperative and defective personality differences evolve, and if this can play a constructive role in the evolution of cooperation?

To gain a better insight into this question, here, I consider a context where individuals possibly play two consecutive PGGs. Individuals have independent strategies in the two PGGs. However, their strategy in the first PGG determines the PGG they enter for the second round. This feature of the model has similarities with stochastic games \cite{Hilbe2018b}, where individuals' strategy can affect the game they play in future rounds. However, while in stochastic games, the future interaction occurs in the same context for all the individuals \cite{Hilbe2018b}, here, I consider a situation where individuals may have different future interactions based on their strategies. I consider two different scenarios. In the first scenario, only first-round cooperators play a second PGG. In the second scenario, both first-round cooperators and first-round defectors are offered the chance to play a second PGG, but in different groups. In the first scenario, playing a second public good can be considered as a potential reward for cooperative behavior. This scenario is motivated by the observations that in many contexts, cooperation serves as a signal of merit \cite{Milinski,Van,Gintis,Salahshour1} or offers a high social status \cite{Smith3,Bird,Marlowe} which can increase others' willingness to interact with a cooperator. This can increase an individual's chance of having future interaction, which I model by allowing a cooperator to enter a second PGG. However, in contrast to a certain reward for a cooperative act, the outcome of the second PGG can be positive or negative, depending on the groups' ability to solve the second public goods dilemma, which we call a prosocial reward dilemma. Then, the question arises whether the community can solve a reward dilemma and if this can promote cooperation in the first PGG in the first place?

In the second scenario, I consider an assortative context where not only first-round cooperators, but both first-round cooperators and first-round defectors enter a second PGG, but in different groups. This scenario is motivated by many pieces of evidence of assortative behavior \cite{Kossinets,Brekke}, such as breaking or forming ties \cite{Wang,Santos2}, according to which individuals are more likely to interact with similar individuals, and can be considered as a context where both cooperative and defective behavior are rewarded, by, respectively, a prosocial and an antisocial reward dilemma. Importantly, in the model introduced here, the strategies of the individuals are independent in the two rounds. Thus, while it is noted that associativity can promote cooperation when individuals have the same strategy in different interactions - a fact which in a sense underlies many mechanisms for the evolution of cooperation such as network reciprocity, tag based mechanisms, group selection and kin selection - it is not clear whether assortativity of the interactions can play a positive role where individuals have a priori independent strategies in different interactions.

The analysis of the models show that cooperation evolves in both the presence of a reward dilemma and in the presence of assortative interaction. As the comparison of the two scenarios reveals, offering defectors the chance to play a second PGG, that is, an antisocial reward dilemma, can have a surprisingly positive impact on the evolution of cooperation. Interestingly, in an assortative context, in the course of evolution, individuals tend to develop consistent strategies in the two consecutive games. By increasing the likelihood that the benefit of a cooperative act is reaped by fellow cooperators, such a personality consistency, in turn, facilitates the evolution of cooperation. These findings shed new light on the evolution of consistent personalities \cite{Weissing,Wolf,Wolf2,Dall,Johnstone,Bergmuller}, and shows how in an attempt to solve social dilemmas, evolution may have given rise to the evolution of consistent cooperative and defective personalities.

\section*{The Model}
In our model, in a well-mixed population of $N$ individuals, groups of size $g$ are randomly formed to play a PGG with enhancement factor $r_1$, possibly followed by a second PGG with enhancement factor $r_2$. That is, at each time step, the whole population is divided into $N/g$ randomly formed groups. Strategies of the individuals in the second PGG are independent of the first PGG. Thus, there are four possible strategies: cooperation in both games ($C_1C_2$), cooperation in the first game and defection in the second one ($C_1D_2$), defection in the first and cooperation in the second game ($D_1C_2$), and finally, defection in both games ($D_1D_2$). I consider two different scenarios. In the first scenario, called a (prosocial) reward dilemma, while defection entails no more round of PGG, cooperation in the first game leads to the entrance to a second PGG. This scenario is consistent with a situation where cooperation serves as a signal of merit \cite{Milinski,Van,Gintis,Salahshour1} or offers a high social status \cite{Smith3,Bird,Marlowe} which can increase others' willingness to interact with a cooperator, and thus, the individual is permitted to enter an elite PGG. In the second scenario, called the assortative public goods game, all the individuals in the group proceed to play a second PGG. However, motivated by many pieces of evidence of assortative behavior \cite{Kossinets,Brekke}, such as breaking or forming ties \cite{Wang,Santos2}, I assume individuals are sorted based on their strategies in the first round, such that all the individuals who cooperate in the first round form a subgroup to play a PGG (which is called the cooperative or prosocial PGG), and all those who defect in the first round form a different subgroup to play their second PGG (which is called the defective or anti-social PGG). In this way, the corresponding PGG can be considered an assortative PGG, in which individuals are sorted based on their strategy in the first round. 

Individuals gather payoff according to the outcome of the games. Besides, I assume individuals receive a base payoff $\pi_0$ from other activities not related to the PGG. After playing the games, individuals reproduce with a probability proportional to their payoff, such that the population size remains constant. That is, the whole population is updated synchronously such that each individual in the next generation is offspring to an individual $i$ in the past generation with a probability proportional to its payoff $\pi_i$. Offspring inherit the strategy of their parent. However, mutations can occur. Mutations in the strategy of the individuals in each round ($s_1$ and $s_2$, where $s_i$ can be $C$ or $D$) occur independently, and each with probability $\nu$. When a mutation occurs, the corresponding variable's value switches to its opposite value ($C$ to $D$ and vice versa).

\begin{figure}%[!hbt]
\centering
\includegraphics[width=1\linewidth, trim = 80 101 55 51, clip,]{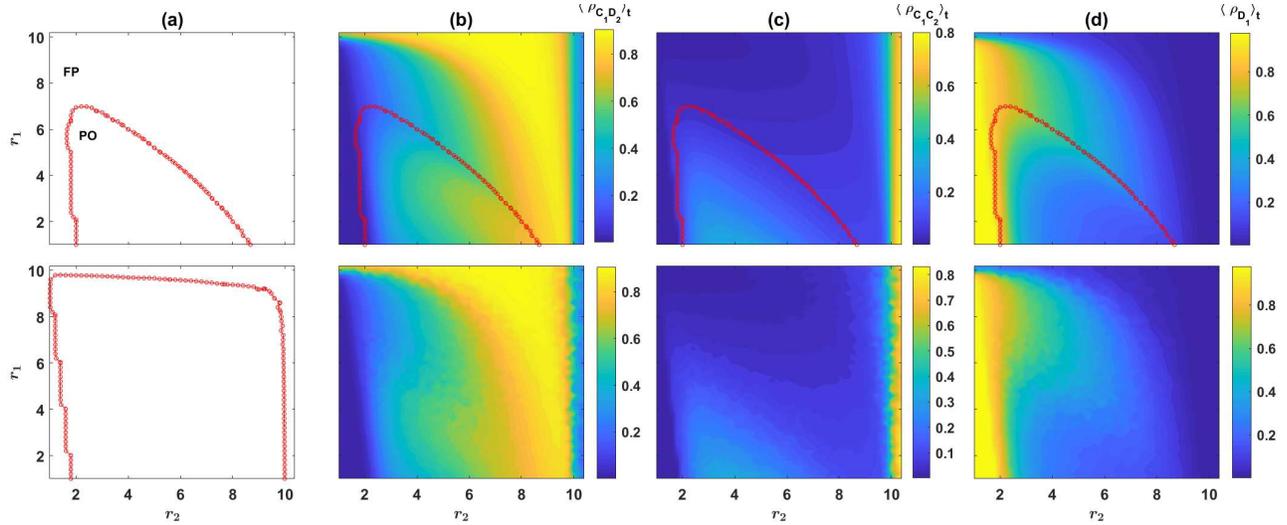}
\caption{A reward dilemma solves the social dilemma. (a): The phase diagram of the model derived from solutions of the replicator dynamics, for two different mutation rates (top $\nu=10^{-3}$ and bottom $\nu=10^{-5}$). The dynamics settle in a periodic orbit or a fixed point depending on the parameters of the model. (b) to (d): Color plots of, respectively, $\langle\rho_{C_1C_2}\rangle_t$, $\langle\rho_{C_1D_2}\rangle_t$ and $\langle\rho_{D_1}\rangle_t=\langle\rho_{D_1C_1}+\rho_{D_1D_2}\rangle_t$. Top panels result from solutions of the replicator dynamics, and bottom panels result from simulations. Cooperation evolves in the second game (c) and is maximized for moderate values of $r_2$. This renders entering the second game an incentive to cooperate in the first game, which promotes cooperation in the first game. Here, $g=10$, $c=1$, and $\pi_0=2$. In (b), (c), and (d), $\nu=10^{-3}$. Simulations are performed in a population of size $N=5000$. The simulation is performed for $T=3000$ time steps starting from an initial condition with random assignment of strategies, and the time averages are taken over the last $2000$ time steps. The replicator dynamics are solved for $T=5000$ time steps starting from homogeneous initial conditions. Time averages are taken over the last $2000$ time steps. As the model is mono-stable in the entire phase diagrams, the results are independent of the initial conditions.}
\label{fig1}
\end{figure}

\section*{Results}
\subsection*{The first scenario: A reward dilemma solves the social dilemma}
As shown in the Methods Section, both models can be described in terms of the replicator dynamics, which gives an exact description of the model in infinite population limit. Beginning with the reward dilemma model, I note that when $r_2>g$, the second PGG is no longer a dilemma. Cooperation becomes the most rational strategy, which guarantees for the second PGG to yield a positive reward to those who enter it by cooperating in the first PGG. As it is known \cite{Attila,Rand,Hilbe}, and as our model confirms, such a certain reward promotes cooperation. The situation becomes more interesting when $r_2<g$. In this case, the second PGG can even yield a negative outcome. For a promise of having future interaction to promote cooperation, the community needs to solve a second dilemma. As I show below, such coupled dilemmas can be solved through evolution. 

I begin by plotting the phase diagram of the model for two different mutation rates, $\nu=10^{-3}$ (top), and $\nu=10^{-5}$ (bottom), in Fig. \ref{fig1}(a). Here and in the following, $g=10$, $c=1$, $\pi_0=2$. For too large values of $r_1$, the dynamics settle in a fixed point (denoted by FP in the figures). On the other hand, for smaller values of $r_1$, cyclic behavior occurs for intermediate values of $r_2$ (indicated by PO in the figures). For both too large and too small $r_2$, a transition to a phase where the dynamics settle in a fixed point is observed. As can be seen by comparing the phase diagram for two different mutation rates, lower mutation rates increase the size of the region where the dynamics settle in a periodic orbit.

To see how the cooperation changes with $r_1$ and $r_2$, in Figs. \ref{fig1}(b), \ref{fig1}(c), and \ref{fig1}(d), I plot, respectively, the time average of $\rho_{C_1C_2}$, $\rho_{C_1D_2}$ and $\rho_{D_1}=\rho_{D_1C_2}+\rho_{D_1D_2}$. For small $r_2$, such that the second PGG is not profitable enough to motivate cooperation in the first PGG, the dynamics settle into a defective fixed point, where the majority of the individuals defect in the first game and do not play the second game. Consequently $\rho_{D_1}$ takes a large value approximately equal to $1-\nu$, and both $\rho_{C_1C_2}$ and $\rho_{C_1D_2}$ take small values maintained by mutations. As $r_2$ increases, cooperation in both rounds evolve. To see how this happens, I note that the cost of cooperation in the first round is equal to $c(r_1/g-1)$. As a single mutant $C_1C_2$ receives a payoff of $c(r_2-1)$ from the second round, cooperation evolves when the payoff of a mutant $C_1C_2$ from the second round becomes larger than the cost of cooperation in the first round. That is when $r_2>2-r_1/g$. By increasing $r_2$ beyond this point, $\rho_{C_1C_2}$ rapidly increases. This increases the effective group size in the second round PGG, and thus the expected payoff of second-round defectors, which is an increasing function in $\rho_{C_1C_2}$, increases. To see why this is the case, I note that the payoff of a $C_1D_2$ individual from the second round in a group composed of $n_{C_1C_2}$ $C_1C_2$ group-mates and $n_{C_1D_2}$ $C_1D_2$ group-mates is equal to $cr_2\rho_{C_1C_2}/(1+n_{C_1C_2}+n_{C_1D_2})$, which is larger in groups with a higher number of $C_1C_2$ individuals. The probability of formation of groups with a higher number of $C_1C_2$ individuals, in turn, increases with increasing $\rho_{C_1C_2}$ (see Methods), and thus, the expected payoff of a second-round defector increases by increasing $\rho_{C_1C_2}$. Consequently $\rho_{C_1D_2}$ increases by further increasing $r_2$ beyond $2-r_1/g$. While this leads to enhanced cooperation in the first round, it also reduces the frequency of second-round cooperators due to the exploitation by second-round defectors. Consequently, the profitability of the second-round PGG decreases and fewer individuals cooperate in the first round to enter the second round PGG. Thus the density of first-round defectors shows a local maximum at a moderate value of $r_2$ at the transition between the cyclic orbit and partially cooperative fixed point at large $r_2$.

\begin{figure}%[!hbt]
\centering
\includegraphics[width=1\linewidth, trim = 90 265 75 05, clip,]{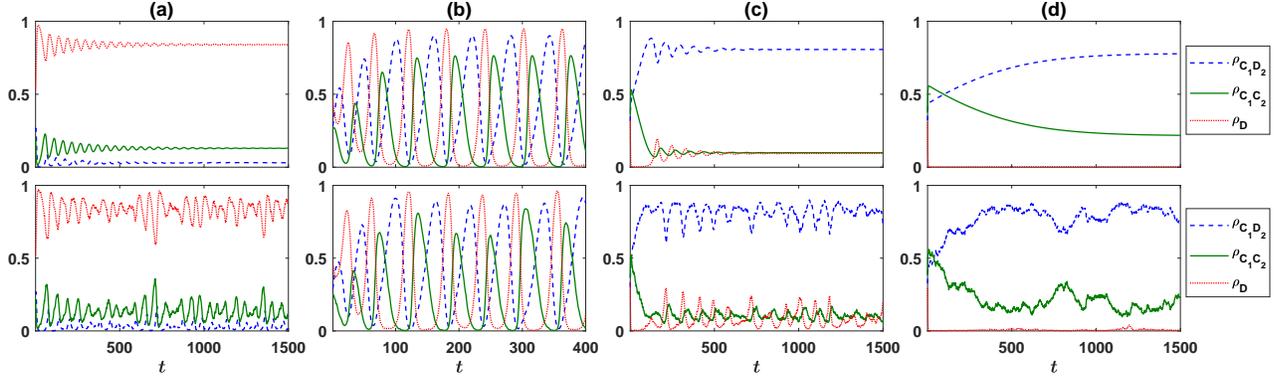}
\caption{Time evolution of the reward dilemma model. The frequency of different strategies as a function of time is plotted. The top panels represent numerical solutions of the replicator dynamics, and the bottom panel results from a simulation in a population of $N=5000$ individuals. In all the cases, $r_1=1.8$. In (a), $r_2=2$, which is slightly larger than the threshold necessary for the evolution of cooperation, $2-r_1/g=1.82$, and thus a small fraction of $C_1C_2$ strategies evolve. In (b) $r_2=3.8$, corresponding to the cyclic phase where different strategies cyclically dominate the population.  In (c) and (d), respectively, $r_2=8.8$ and $r_2=9.8$, both corresponding to the cooperative fixed point. Parameter values: $\nu=10^{-3}$, $g=10$, $c=1$, and $\pi_0=2$. The replicator dynamics is solved starting from homogeneous initial conditions, and simulations are performed starting from random assignment of strategies.}
\label{fig2}
\end{figure}

The time evolution of the system for different strength of reward dilemma $r_2$ is presented in Fig. \ref{fig2}. Here, the result of the numerical solution of the replicator dynamics (top) and a simulation in a population of size $N=5000$ (bottom) are presented. Here, $g=10$, $\nu=10^{-3}$, $c=1$, $\pi_0=2$, and $r_1=1.8$. In Fig. \ref{fig2}(a), $r_2=2$. This is slightly larger than $2-r_1/g=1.82$, and thus $C_1C_2$ strategy survive in the population. For larger values of $r_2$, the fixed point becomes unstable, and the dynamics settle in the cyclic orbit. An example of the cyclic orbit for $r_2=3.8$ is presented in Fig. \ref{fig2}(b). In the cyclic phase, when the density of individuals who cooperate in the first round and thus enter the second PGG is small, individuals can reach a high payoff by entering and cooperating in the second PGG. Thus $\rho_{C_1C_2}$ increases. When $\rho_{C_1C_2}$ increases enough, individuals can reach a higher payoff by defecting in the second PGG. At this point, $\rho_{C_1D_2}$ begins to increase, while $\rho_{C_1C_2}$ decreases. As the density of defectors in the second PGG increases, its profitability decreases, and thus individuals have no incentive to cooperate in the first round. Consequently, both $\rho_{D_1D_2}$ and $\rho_{D_1C_2}$, as well as $\rho_D=\rho_{D_1D_2}+\rho_{D_1C_2}$ increase, while other strategies decrease (I note that, since those who defect in the first round do not enter the second game, the two strategies $D_1D_2$ and $D_1C_2$ are degenerate as they lead to the same payoff and are found in the same densities).

The time evolution of the system in the partially cooperative fixed point in large $r_2$ is presented in Figs. \ref{fig2}(c) and \ref{fig2}(d). In Fig. \ref{fig2}(c), $r_2=8.8$. This corresponds to just above the transition line from the periodic solution to the fixed point. Consequently, the replicator dynamics settle in the fixed point after showing transient damped osculations around the fixed point. The simulation results show small fluctuations around the stationary state due to finite-size effects. Comparison of the case of $r_2=3.8$ in Fig. \ref{fig2}(b) and $r_2=8.8$ in Fig. \ref{fig2}(c) shows that $\rho_{C_1D_2}$ increases for larger $r_2$ due to the higher profitability of the second-round PGG, which motivates more individuals to cooperate in the first round to enter this PGG. This, in turn, can have an adverse effect on $\rho_{C_1C_2}$ due to the larger effective group size of the second round $PGG$. For larger $r_2$, as in Fig. \ref{fig2}(d) where $r_2=9.8$, the dynamics settles in the fixed point without showing damped osculations. Furthermore, $\rho_{C_1C_2}$ increases and $\rho_{C_1D_2}$ decrease as $r_2$ approaches $g$.

\begin{figure}%[!hbt]
\centering
\includegraphics[width=1\linewidth, trim = 196 255 280 90, clip,]{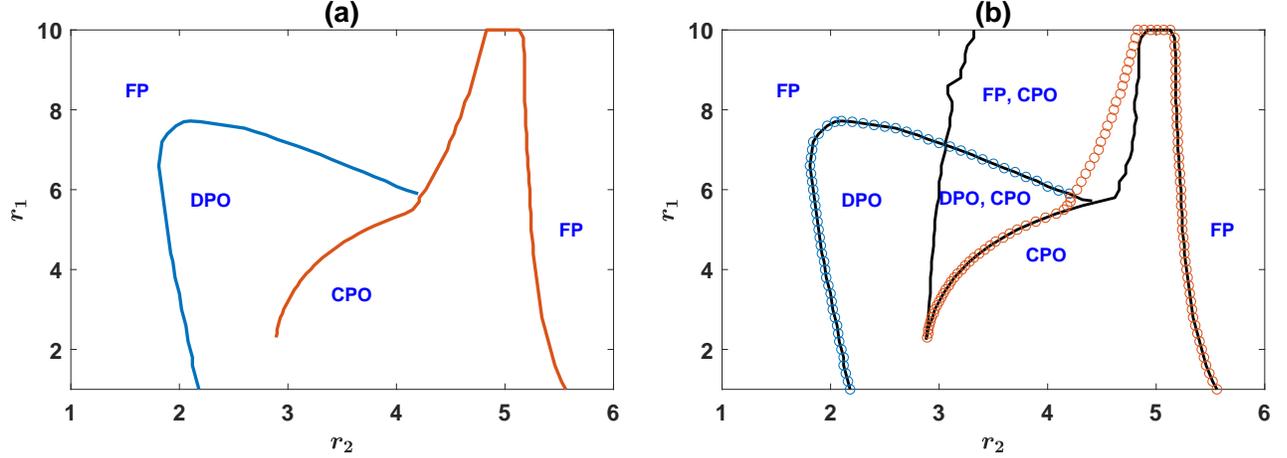}
\caption{The phase diagram of the assortative public goods game. The phase diagram in the $r_1-r_2$ plane. For both small and large values of $r_2$ the model settles into a fixed point (FP). In between, the model shows the cyclic dominance of different strategies. Two different periodic orbits, DPO and CPO, separated by a singular transition exist. (b): Different stability regions in the $r_1-r_2$ plane (lines). The phase boundaries are superimposed (empty circles). The model shows two bistability regions where either both the fixed point and the CPO, or both DPO and CPO are stable. Here, $g=10$, $\nu=10^{-3}$, $c=1$, and $\pi_0=2$.}
\label{fig3}
\end{figure}

\subsection*{The second scenario: The evolution of cooperation and consistent personalities in assortative public goods}
As we have seen so far, a reward dilemma solves the social dilemma. An interesting question is whether such a mechanism can be competitive if defectors have the chance of forming a PGG of their own? This brings us to an assortative public goods game where both first-round cooperators and first-round defectors are rewarded by a second round of interaction, but in separate groups.

The phase diagram of the assortative public goods game is presented in Fig(\ref{fig3}.a), top panel. For both too small and too large $r_2$, the system settles into a fixed point, and cyclic behavior emerges in between. However, there are two qualitatively different periodic orbits, each stable in some region of the parameter space. For smaller $r_2$, the dynamics settle into a defective periodic orbit (DPO). In this orbit, while cooperation in the cooperative PGG evolves, cooperation does not evolve in the defective PGG. I note that this is the same periodic orbit observed in the reward dilemma model. That such a periodic orbit endures in the second scenario shows its competitive stability. In other words, just as prosocial reward is stable in the presence of an antisocial reward \cite{Attila}, a prosocial reward dilemma is stable in the presence of an anti-social reward dilemma. On the other hand, for large $r_2$, the dynamics settle into the cooperative periodic orbit (CPO), where cooperation in both cooperative and defective PGGs evolves.

\begin{figure}%[ht]
\centering
\includegraphics[width=1\linewidth, trim = 90 265 75 05, clip,]{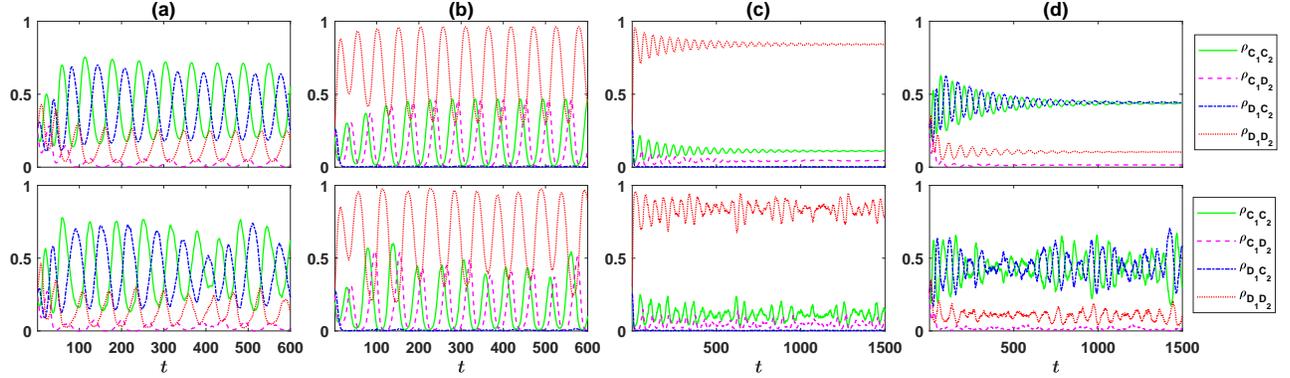}
\caption{The time evolution of the system in assortative public goods games (a) to (d): The frequency of different strategies as a function of time. Top panels show replicator dynamics solutions and the bottom panels result form the simulations in a population of $N=5000$ individuals. The system shows two different periodic orbits. A cooperative periodic orbit, with higher level of cooperation for larger $r_2$, $r_2=4.8$ in (a), and a defective periodic orbit, for small $r_2$, $r_2=2.4$ in (b). In the fixed point of the dynamics for small $r_2$ but larger than $2-r_1/g$, as in (c) where $r_2=2$, only $C_1C_2$ cooperators evolve, and for large $r_2$, $r_2=5.6$ in (d), both $C_1C_2$ and $D_1C_2$ evolve. Here, $g=10$, $\nu=10^{-3}$, $c=1$, $\pi_0=2$, and $r_1=2.8$. The replicator dynamics is solved starting from homogeneous initial conditions and simulations are performed starting from random assignment of strategies.}
\label{fig4}
\end{figure}

Interestingly, for small fixed $r_1$, by increasing $r_2$ the system shows a cross-over from the DPO to CPO without passing any singularity. However, for large $r_1$, starting from a uniform initial condition, in which the initial densities of all the strategies are equal, as $r_2$ increases, the equilibrium state of the system changes singularly in a certain value of $r_2$. This indicates the transition between DPO and CPO resembles a discontinuous transition for larger $r_1$. As a discontinuous transition is usually accompanied by bistability \cite{Binder}, this raises the question of whether the system possesses a bistable region as well? To address this question, I present the boundaries of bistability in Fig. \ref{fig3}(b) (black lines). The phase boundaries, which result from a homogeneous initial condition, are superimposed in this figure as well. The boundaries of bistability are derived by solving the replicator dynamics starting from different initial conditions and checking for hysteresis (see Methods). The system is monostable outside of the bistable region, indicated by black lines. In the monostable regions, the dynamics settle in the same stationary state, starting from all the initial conditions. In the bistable region (inside the black line), two different stable states, indicated in the figure, are possible depending on the initial conditions. I note that, while the cooperative periodic orbit is stable in the bistable region, it has a very small basin of attraction, such that the replicator dynamics does not settle into this orbit starting from most randomly generated initial conditions (see the Supplementary Information, figures SI.4 and SI.5). For this reason, to derive the boundaries of bistability, a hysteresis analysis is used (see methods).

As can be seen in the figure, the transition from a fixed point to the DPO in small $r_2$ does not show any bistability and occurs at the same value for all the initial conditions. Similarly, the transition from CPO to a fixed point in large $r_2$ does not show any bistability. In contrast, the transition to CPO by increasing $r_2$ shows bistability: For medium $r_2$, there is a region of the phase diagram where both CPO and DPO are stable. Similarly, the model shows a bistability region for large $r_1$ where both the fixed point and the cooperative periodic orbit are stable. 

In Fig. \ref{fig4}, I present the time evolution of different strategies. An example of different periodic orbits is presented in Fig. \ref{fig4}(a) (CPO) and Fig. \ref{fig4}(b) (DPO). Top panels present the result of the replicator dynamics, and the bottom panels present the result of a simulation in a population of size $N=5000$. In Fig. \ref{fig4}(a), $r_1=2.8$ and $r_2=4.8$, and in Fig. \ref{fig4}(b), $r_1=2.8$ and $r_2=2.4$. For larger $r_2$, as in Fig. \ref{fig4}(a), defective and cooperative PGGs perform competitively, and competition between these two maintains cooperation in the system. When $\rho_{C_1D_1}$ ($\rho_{D_1D_2}$) is small, the cooperative (defective) PGG is profitable, and thus, it motivates individuals to cooperate (defect) in the first game in order to enter to this PGG. Consequently, $\rho_{C_1C_2}$ ($\rho_{D_1C_2}$) increases. As $\rho_{C_1C_2}$ ($\rho_{D_1C_2}$) increases enough, the cooperative (defective) PGG becomes vulnerable to defection, due to high frequency of cooperators in this PGG which increase the expected payoff of a second round defector, $\rho_{C_1D_2}$ ($\rho_{D_1D_2}$). At this point, $\rho_{C_1D_1}$ ($\rho_{D_1D_2}$) starts to increase. This in turn decreases the profitability of the cooperative (defective) PGG, and individuals are better off by switching to defection (cooperation) in the first round to enter the defective (cooperative) PGG. Consequently, both $\rho_{C_1C_2}$ and $\rho_{C_1D_2}$ ($\rho_{D_1C_2}$ and $\rho_{D_1D_2}$) decrease, while $\rho_{D_1C_2}$ and $\rho_{D_1D_2}$ ($\rho_{C_1C_2}$ and $\rho_{C_1D_2}$) increase. In this way, competition between cooperative and defective PGGs maintain cooperation in the population. Interestingly, individuals tend to have compatible strategies in the two rounds. That is, those who cooperate in the first round are more likely to cooperate in the second round. This can be seen by noting that on average $\rho_{C_1D_2}$ is much smaller than $\rho_{D_1D_2}$, even though defection in cooperative and defective PGGs leads to, on average, similar payoffs as the density of cooperators in these two games ($\rho_{C_1C_2}$ and $\rho_{D_1C_2}$) are similar.

\begin{figure}[!hbt]
\centering
\includegraphics[width=1\linewidth, trim = 81 51 40 95, clip,]{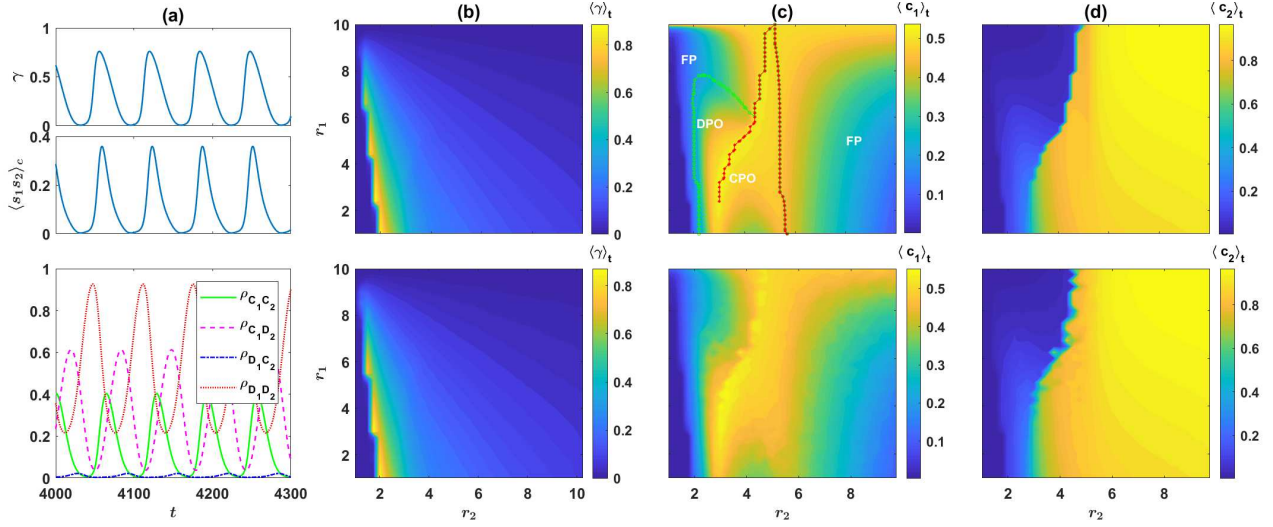}
\caption{Evolution of cooperation and consistent personalities in assortative public goods game. (a): The time series of the personality consistency measure $\gamma$ (top), and the connected correlation function between the strategies of the individuals in the two rounds $\langle s_1s_2\rangle_t$ (middle). For comparison, the densities of different strategies are plotted in the bottom panel. Both measures always remain non-negative. This shows individuals evolve consistent personalities in the two rounds. (b): The contour plot of the time average personality measure $\langle\gamma\rangle_t$, in $r_1-r_2$ plane. $\gamma$ remains non-negative in the whole phase diagram, which indicates consistent cooperative and defective personalities evolve. (c): The contour plot of the time average cooperation level in the first round $\langle\rho_{C_1}\rangle_t=\langle\rho_{C_1C_2}+\rho_{C_1D_2}\rangle_t$. The phase boundaries are plotted in the top panel as well. Interestingly, the cooperation level in the first round is maximized on the singular transition between the two periodic orbits. (e): The contour plot of the time average cooperation level in the second round $\langle\rho_{C_2}\rangle_t=\langle\rho_{C_1C_2}+\rho_{D_1C_2}\rangle_t$. $\langle\rho_{C_2}\rangle_t$ increases with increasing $r_2$. Here, $g=10$, $\nu=10^{-3}$, $c=1$, and $\pi_0=2$. In (a) numerical solutions of the replicator dynamics are used. In (b) to (d), top panels result from numerical solutions of the replicator dynamics, and bottom panels result from simulation on a population of size $N=10000$. The replicator equations are solved for $T=5000$ time steps starting from a homogeneous initial condition, and the time averages are taken over the last $2000$ steps. The simulations are performed for $T=4000$ steps starting from a random assignment of strategies, and the averages are taken over the last $3500$ steps.}
\label{fig5}
\end{figure}

As mentioned before, the situation is different for smaller $r_2$. For smaller $r_2$, as can be seen in Fig. \ref{fig4}(b), while cooperation in the cooperative PGG evolves, cooperation in the defective PGG does not evolve. This again hints at the evolution of compatible strategies in the two rounds. Consequently, for smaller $r_2$, while the cooperative PGG can become profitable due to the evolution of fully cooperative $C_1C_2$ strategies and motivates individuals to cooperate in the first game, the defective PGG does not perform competitively and can not attract individuals. As mentioned before, this periodic orbit is the same periodic orbit observed in the reward dilemma model.

The dynamics settle in fixed points for both small and large $r_2$. Example of the dynamics of the system in this regime are presented in Fig. \ref{fig4}(c), for small $r_2$, $r_2=2$ and Fig. \ref{fig4}(d), for large $r_2$, $r_2=5.6$. As for small $r_2$, cooperation evolves only in the cooperative PGG, the situation is similar to the reward dilemma model: $C_1C_2$ individuals are found in the population (beyond that maintained by mutations) if $r_2>2-r_1/g$. This is what we observe in Fig. \ref{fig4}(c). This again hints at the evolution of consistent personalities, as only fist round cooperators cooperate in the second round. The fixed point for large $r_2$ is presented in Fig. \ref{fig4}(d). As $r_2$ is chosen just slightly beyond the transition line between cooperative periodic orbit and the fixed point, the dynamics shows damped oscillations around the stationary state before settling in the fixed point. Simulations in a finite population, on the other hand, show that small fluctuations around the stationary state occur in this region. I note that the evolution of consistent personalities can be observed in this regime as well, as the frequency of defectors in the defective PGG is larger than that in the cooperative PGG. That is, first-round defectors are more likely to defect in the second round compared to first-round cooperators.

The fact that individuals' strategies in the two games tend to be compatible can be studied in more depth. To do this, I consider two measures of consistency of the strategies in the two rounds. As a first measure of compatibility of the strategies of the individuals in the two rounds, I define $\gamma=[P(C_2|C_1)+P(D_2|D_1)-P(D_2|C_1)-P(C_2|D_1)]/2$. Here, $P(s_2|s_1)$ is the conditional probability that an individual has strategy $s_2$ in the second game, given its strategy in the first game $s_1$. $\gamma$ takes a value between one and minus one, and the more positive $\gamma$, the more individuals' strategies in the two rounds are consistent. As a second measure of personality consistency, I consider the connected correlation function of the strategies of the individuals in the two rounds $\langle s_1 s_2 \rangle_c= \langle s_1s_2\rangle-\langle s_1\rangle \langle s_2\rangle$. To calculate this, I assign $-1$ to the strategy $D$, and $+1$ to the strategy $C$. These are plotted in Fig. \ref{fig5}(a) (top). For comparison, the density of different strategies is plotted as well (bottom panel). Both personality measures show cyclic behavior and always remain non-negative. Importantly, the latter holds in all the phases. This can be seen in Fig. \ref{fig5}(b), where the color plot of the time average of $\gamma$, for numerical solutions of the replicator dynamics (top) and a simulation in a population of size $N=10000$ (bottom) is presented. Interestingly, for fixed $r_2$, personality consistency measures show a maximum close to (but not exactly on) the transition from the fixed point to the defective periodic orbit. This corresponds to $r_2=2-r_1/g$, where the benefit of cooperation in the cooperative PGG starts to become large enough to compensate the cost of cooperation in the first round necessary to enter this PGG. Our results thus show that individuals develop consistent cooperative and defective personalities [See the Supplementary Information for the validity of this result for other parameter values]. This, in turn, plays a positive role in promoting cooperation, as individuals behaving consistently in the two rounds, together with the assortative nature of the public goods, allows first-round cooperators to be more likely to reap the benefit of cooperation in the second round, compared to first-round defectors.  

I begin the study of the cooperation by plotting the average cooperation in the first game in Fig. \ref{fig5}(c), and the average cooperation in the second game in Fig. \ref{fig5}(d). Here, the phase transition lines are indicated in the figure as well. As can be seen, cooperation in the first game is maximized for a moderate value of $r_2$, and it drops for both too small and too large values of $r_2$. Increasing $r_2$ beyond this value has a detrimental effect on cooperation in the first round. This is due to the fact that for larger values of $r_2$, cooperation in the second round increases in both the defective and the cooperative PGGs. This decreases individuals' incentive to cooperate in the first round to be sorted with fellow cooperators in the second round. On the other hand, for too small $r_2$, any potential benefit from the second round can be too small to promote high cooperation in the first round. Interestingly, the maximum cooperation level is achieved exactly on the transition line between the cooperative and the defective periodic orbits, which coincides with the edge of bistability. This aligns with some arguments that being on the edge of bistability can be beneficial for biological systems \cite{Salahshour}.

Finally, the level of cooperation in the second PGG is plotted in Fig. \ref{fig5}(d). Here, it can be seen the cooperation increases by increasing $r_2$. On the other hand, as we have seen in Fig. \ref{fig1}, in the reward dilemma model, cooperation in the second game is maximized in an intermediate value of $r_2$. This contrast can be argued to result from competition between the two cooperative and defective PGGs, such that evolution favors the more cooperative one. Consequently, if individuals in one of the two cooperative or defective PGGs start to defect, the other more cooperative one is favored by evolution. This, in turn, shows that a potential reward to defection, by promoting competition, can have a surprisingly positive impact on the evolution of cooperation in an assortative context.

\section*{Discussion}
The fact that individuals consistently show cooperative or defective strategies had been noticed in public goods experiments \cite{Fischbacher1,Fischbacher2,Burlando,Chaudhuri}, and in many animal populations \cite{Bergmuller}. It is argued this persistent personality differences can partly explain why cooperation is observed in laboratory experiments and many human and animal societies \cite{Bergmuller,Fischbacher1,Fischbacher2,Burlando,Chaudhuri}. However, while some theoretical work had shed light on the evolution of some aspects of personality differences, such as the evolution of responsive and unresponsive personalities\cite{Wolf2}, risk-averse and risk-taking personalities \cite{Weissing}, or personality differences in leadership \cite{Johnstone}, the evolution of cooperative and defective personalities had alluded theoretical understanding. Our findings address this gap by showing how such consistent personality differences can evolve naturally in an evolutionary process when individuals need to work collectively to solve a social dilemma. Importantly, as the analysis of the model shows, the evolution of consistent personalities, in turn, helps solving social dilemmas by increasing the likelihood that the benefit of a cooperative act is reaped by those who behave cooperatively. In this regard, the evolution and maintenance of consistent cooperative and defective personalities can be regarded as an important mechanism at work in promoting cooperation in biological populations.

An interesting question is that why consistent personality differences evolve in an assortative context? The key to the question is that entering the cooperative PGG in the second round is costly, while entering the defective PGG is costless. As it has been shown recently, an entrance cost for a PGG can promote cooperation in a costly PGG, due to the smaller effective size of a costly PGG \cite{Salahshour2021A}. For this reason, cooperation is more frequent and defection less frequent in the cooperative, costly PGG, compared to the defective, cost-less PGG. This phenomenon naturally leads to the evolution of consistent cooperative and defective personalities in an assortative context.

Our analysis also reveals new roads to the evolution of cooperation. In this regard, as the analysis of the reward dilemma model shows, a population of self-interested agents can successfully solve a reward dilemma and this, in turn, helps to solve the social dilemma. This mechanism can be at work to promote cooperation in a context where cooperation increases an individual's chance of having more social interaction, even if actually benefiting from those interactions requires solving another social dilemma. The second scenario model, on the other hand, shows cooperation still evolves when defection is rewarded by a promise of future interaction as well, provided an assortative mechanism is at work. In other words, just as prosocial reward is stable against antisocial reward \cite{Attila}, a prosocial reward dilemma is competitively stable in promoting cooperation in the presence of an anti-social reward dilemma. In the presence of both prosocial and antisocial reward dilemmas, competition between the prosocial and anti-social public goods maintains cooperation in the system, and moreover, surprisingly increase cooperation in the second round, compared to a case where such competition is lacking. This shows a potential reward to defection, by fostering competition, can have a surprisingly positive impact in promoting cooperation.

\section*{Methods}
\label{Methods}
\subsection*{Replicator dynamics}
The model can be described in terms of replicator-mutation equations \cite{Nowak3}, which provide an exact description of the model in infinite population limit. These equations can be written as follows:
\begin{align}
\rho_{xy}(t+1)=\sum_{x',y'}\nu_{xy}^{x'y'}\rho_{x'y'}(t)\frac{\pi_{x'y'}(t)}{\sum_{x'',y''}\rho_{x''y''}(t)\pi_{x''y''}(t)}.
\label{eqrepMmain}
\end{align}
Here, $xy$ (as well as $x'y'$ and $x''y''$) refer to the strategies of the individuals, such that $x$ is the strategy of an individual in the first round, and $y$ is its strategy in the second round. $x$, $x'$ etc. can be either cooperation $C$ or defection $D$. $\nu_{xy}^{x'y'}$ is the mutation rate  from the strategy $x'y'$ to the strategy $xy$. These can be written in terms of mutation rate $\nu$ as follows: 
\begin{align}
\begin{cases}
\nu_{xy}^{x'y'}=1-2\nu+\nu^2 \quad &\textit{if\quad  ($x=x'$ \quad and \quad $y=y'$)},\\ 
\nu_{xy}^{x'y'}=\nu-\nu^2 \quad&\textit{if\quad ($x\neq x'$ \quad and \quad $y=y'$)\quad  or \quad ($x=x'$ \quad and \quad $y\neq y'$)}, \\
\nu_{xy}^{x'y'}=\nu^2\quad& \textit{if\quad ($x\neq x'$\quad and \quad$y \neq y'$)}.
\end{cases}
\end{align}
In eq. (\ref{eqrepMmain}), $\pi_{x'y'}$ is the expected payoff of an individuals with strategy $x'y'$. In the case of the first scenario, these are given by the following equations:
\begin{align}
\pi_{C_1C_2}=&\sum_{n_{D_1C_2}=0}^{g-1-n_{C_1C_2}-n_{C_1D_2}}\sum_{n_{C_1D_2}=0}^{g-1-n_{C_1C_2}}\sum_{n_{C_1C_2}=0}^{g-1}\bigg[cr_1\frac{1+n_{C_1}}{g} +cr_2\frac{1+n_{C_1C_2}}{1+n_{C_1}}\bigg]{\rho_{C_1C_2}}^{n_{C_1C_2}}{\rho_{C_1D_2}}^{n_{C_1D_2}}{\rho_{D_1C_2}}^{n_{D_1C_2}}\nonumber\\&{\rho_{D_1D_2}}^{g-1-n_{C_1C_2}-n_{C_1D_2}-n_{D_1C_2}}\binom{g-1}{n_{C_1C_2},n_{C_1D_2},n_{D_1C_2},g-1-n_{C_1C_2}-n_{C_1D_2}-n_{D_1C_2}}-2c+\pi_0,\nonumber\\
\pi_{C_1D_2}=&\sum_{n_{D_1C_2}=0}^{g-1-n_{C_1C_2}-n_{C_1D_2}}\sum_{n_{C_1D_2}=0}^{g-1-n_{C_1C_2}}\sum_{n_{C_1C_2}=0}^{g-1}\bigg[cr_1\frac{1+n_{C_1}}{g} +cr_2\frac{n_{C_1C_2}}{1+n_{C_1}}\bigg]{\rho_{C_1C_2}}^{n_{C_1C_2}}{\rho_{C_1D_2}}^{n_{C_1D_2}}{\rho_{D_1C_2}}^{n_{D_1C_2}}\nonumber\\&{\rho_{D_1D_2}}^{g-1-n_{C_1C_2}-n_{C_1D_2}-n_{D_1C_2}}\binom{g-1}{n_{C_1C_2},n_{C_1D_2},n_{D_1C_2},g-1-n_{C_1C_2}-n_{C_1D_2}-n_{D_1C_2}}-c+\pi_0,\nonumber\\
\pi_{D_1C_2}=&\sum_{n_{D_1C_2}=0}^{g-1-n_{C_1C_2}-n_{C_1D_2}}\sum_{n_{C_1D_2}=0}^{g-1-n_{C_1C_2}}\sum_{n_{C_1C_2}=0}^{g-1}\bigg[cr_1\frac{n_{C_1}}{g} \bigg]{\rho_{C_1C_2}}^{n_{C_1C_2}}{\rho_{C_1D_2}}^{n_{C_1D_2}}{\rho_{D_1C_2}}^{n_{D_1C_2}}\nonumber\\&{\rho_{D_1D_2}}^{g-1-n_{C_1C_2}-n_{C_1D_2}-n_{D_1C_2}}\binom{g-1}{n_{C_1C_2},n_{C_1D_2},n_{D_1C_2},g-1-n_{C_1C_2}-n_{C_1D_2}-n_{D_1C_2}}+\pi_0,\nonumber\\
\pi_{D_1D_2}=&\sum_{n_{D_1C_2}=0}^{g-1-n_{C_1C_2}-n_{C_1D_2}}\sum_{n_{C_1D_2}=0}^{g-1-n_{C_1C_2}}\sum_{n_{C_1C_2}=0}^{g-1}\bigg[cr_1\frac{n_{C_1}}{g} \bigg]{\rho_{C_1C_2}}^{n_{C_1C_2}}{\rho_{C_1D_2}}^{n_{C_1D_2}}{\rho_{D_1C_2}}^{n_{D_1C_2}}\nonumber\\&{\rho_{D_1D_2}}^{g-1-n_{C_1C_2}-n_{C_1D_2}-n_{D_1C_2}}\binom{g-1}{n_{C_1C_2},n_{C_1D_2},n_{D_1C_2},g-1-n_{C_1C_2}-n_{C_1D_2}-n_{D_1C_2}}+\pi_0.
\label{eqscenario1Mmain}
\end{align}
Here, we have $n_{C_1}=n_{C_1C_2}+n_{C_1D_2}$. To write these equations, I used the fact that in a group with $n_{C_1C_2}$ individuals with strategy ${C_1C_2}$, and $n_{C_1D_2}$ individuals with strategy ${C_1D_2}$, $r_1\frac{1+n_{C_1}}{g}-c$ and $r_1\frac{n_{C_1}}{g}$ are, respectively, the expected payoff of an individual who cooperates, defects, in the first game. Those who defect in the first game do not gather payoff from the second game. On the other hand, those who cooperate in the first game, obtain a payoff from the second game as well (which can be negative or positive). This is $r_2\frac{1+n_{C_1C_2}}{1+n_{C_1}}-c$ for an individual with strategy $C_1C_2$, and $r_2\frac{n_{C_1C_2}}{1+n_{C_1}}$ for an individual with strategy $C_1D_2$. Finally, ${\rho_{C_1C_2}}^{n_{C_1C_2}}{\rho_{C_1D_2}}^{n_{C_1D_2}}{\rho_{D_1C_2}}^{n_{D_1C_2}}$ ${\rho_{D_1D_2}}^{g-1-n_{C_1C_2}-n_{C_1D_2}-n_{D_1C_2}}\binom{g-1}{n_{C_1C_2},n_{C_1D_2},n_{D_1D_2},g-1-n_{C_1C_2}-n_{C_1D_2}-n_{D_1C_2}}$, is the probability that a focal individual finds itself in a group with $n_{C_1C_2}$, $n_{C_1D_2}$, $n_{D_1C_2}$, and $n_{D_1D_2}$ individuals with, respectively, strategies $C_1C_2$, $C_1D_2$, $D_1C_2$, and $D_1D_2$. Here, $\binom{g-1}{n_{C_1C_2},n_{C_1D_2},n_{D_1D_2},g-1-n_{C_1C_2}-n_{C_1D_2}-n_{D_1C_2}}$ is the multinational coefficients (that is the number of ways that among the $g-1$ group mates of a focal individual, $n_{C_1C_2}$, $n_{C_1D_2}$, $n_{D_1C_2}$, $g-1-n_{C_1C_2}-n_{C_1D_2}-n_{D_1C_2}$ individuals have strategies, respectively, $C_1C_2$, $C_1D_2$, $D_1C_2$, and $D_1D_2$). Summation over all the possible configurations gives the expected payoff of the focal individual with the given strategy from the games. Finally, as all the individuals receive a base payoff $\pi_0$, this is added to the total payoff. Using the expressions in eq. (\ref{eqscenario1Mmain}) for the expected payoff of different strategies in eq. (\ref{eqrepMmain}), I have a set of four equations which gives an analytical description of the model, in the limit of infinite population size. 

In the same way, it is possible to write down equations for the expected payoffs of individuals with different strategies in the second scenario. The difference with the preceding scenario is that, in the second scenario those who defect in the first round proceed to a second PGG as well. Thus, under the same notation and conventions as before, the individuals with strategies $D_1C_2$ and $D_1D_2$, obtain a payoff of, respectively, $r_2\frac{1+n_{D_1C_2}}{1+n_{D_1}}-c$ and $r_2\frac{n_{D_1C_2}}{1+n_{D_1}}$, from  their second game. Here, $n_{D_1}=n_{D_1C_1}+n_{D_1D_2}$. Thus, we have for the expected payoffs of different strategies in the second scenario:
\begin{align}
\pi_{C_1C_2}=&\sum_{n_{D_1C_2}=0}^{g-1-n_{C_1C_2}-n_{C_1D_2}}\sum_{n_{C_1D_2}=0}^{g-1-n_{C_1C_2}}\sum_{n_{C_1C_2}=0}^{g-1}\bigg[cr_1\frac{1+n_{C_1}}{g} +cr_2\frac{1+n_{C_1C_2}}{1+n_{C_1}}\bigg]{\rho_{C_1C_2}}^{n_{C_1C_2}}{\rho_{C_1D_2}}^{n_{C_1D_2}}{\rho_{D_1C_2}}^{n_{D_1C_2}}\nonumber\\&{\rho_{D_1D_2}}^{g-1-n_{C_1C_2}-n_{C_1D_2}-n_{D_1C_2}}\binom{g-1}{n_{C_1C_2},n_{C_1D_2},n_{D_1C_2},g-1-n_{C_1C_2}-n_{C_1D_2}-n_{D_1C_2}}-2c+\pi_0,\nonumber\\
\pi_{C_1D_2}=&\sum_{n_{D_1C_2}=0}^{g-1-n_{C_1C_2}-n_{C_1D_2}}\sum_{n_{C_1D_2}=0}^{g-1-n_{C_1C_2}}\sum_{n_{C_1C_2}=0}^{g-1}\bigg[cr_1\frac{1+n_{C_1}}{g} +cr_2\frac{n_{C_1C_2}}{1+n_{C_1}}\bigg]{\rho_{C_1C_2}}^{n_{C_1C_2}}{\rho_{C_1D_2}}^{n_{C_1D_2}}{\rho_{D_1C_2}}^{n_{D_1C_2}}\nonumber\\&{\rho_{D_1D_2}}^{g-1-n_{C_1C_2}-n_{C_1D_2}-n_{D_1C_2}}\binom{g-1}{n_{C_1C_2},n_{C_1D_2},n_{D_1C_2},g-1-n_{C_1C_2}-n_{C_1D_2}-n_{D_1C_2}}-c+\pi_0,\nonumber\\
\pi_{D_1C_2}=&\sum_{n_{D_1C_2}=0}^{g-1-n_{C_1C_2}-n_{C_1D_2}}\sum_{n_{C_1D_2}=0}^{g-1-n_{C_1C_2}}\sum_{n_{C_1C_2}=0}^{g-1}\bigg[cr_1\frac{n_{C_1}}{g} +cr_2\frac{1+n_{D_1C_2}}{1+n_{D_1}}\bigg]{\rho_{C_1C_2}}^{n_{C_1C_2}}{\rho_{C_1D_2}}^{n_{C_1D_2}}{\rho_{D_1C_2}}^{n_{D_1C_2}}\nonumber\\&{\rho_{D_1D_2}}^{g-1-n_{C_1C_2}-n_{C_1D_2}-n_{D_1C_2}}\binom{g-1}{n_{C_1C_2},n_{C_1D_2},n_{D_1C_2},g-1-n_{C_1C_2}-n_{C_1D_2}-n_{D_1C_2}}-c+\pi_0,\nonumber
\end{align}
\begin{align}
\pi_{D_1D_2}=&\sum_{n_{D_1C_2}=0}^{g-1-n_{C_1C_2}-n_{C_1D_2}}\sum_{n_{C_1D_2}=0}^{g-1-n_{C_1C_2}}\sum_{n_{C_1C_2}=0}^{g-1}\bigg[cr_1\frac{n_{C_1}}{g} +cr_2\frac{n_{D_1C_2}}{1+n_{D_1}}\bigg]{\rho_{C_1C_2}}^{n_{C_1C_2}}{\rho_{C_1D_2}}^{n_{C_1D_2}}{\rho_{D_1C_2}}^{n_{D_1C_2}}\nonumber\\&{\rho_{D_1D_2}}^{g-1-n_{C_1C_2}-n_{C_1D_2}-n_{D_1C_2}}\binom{g-1}{n_{C_1C_2},n_{C_1D_2},n_{D_1C_2},g-1-n_{C_1C_2}-n_{C_1D_2}-n_{D_1C_2}}+\pi_0.
\end{align}
Using these expressions for the expected payoffs of individuals with different strategies in eq. (\ref{eqrepMmain}), we have the analytical description of the second scenario model, in the limit of infinite population size.
\subsection*{The simulations and numerical solutions}
Numerical solutions of the replicator dynamics result from numerically solving the replicator dynamics of the models derived in the Methods section. Simulations of the models are performed according to the model definition. Unless otherwise stated, both the simulations and numerical solutions of the replicator dynamics are performed with an initial condition in which all the strategies are found in similar frequencies in the population pool. For the solutions of the replicator dynamics, this is assured by setting the initial frequency of all the four strategies equal to $1/4$. For simulations, this is assured by a random assignment of the strategies. The phase diagram presented in Fig. \ref{fig5}(a) is derived by locating the parameter values where a transition between different attractors occurs starting from a homogeneous initial condition. The boundary of bistability in Fig. \ref{fig5}(a) is derived by examining history dependence and checking for the existence of hysteresis in the evolution of the system. That is, the replicator dynamics are solved starting from parameter values belonging to different phases. Then, the parameter values are changed in small steps, using the stationary state of the preceding steps as the initial condition for the solution of the replicator dynamics in the next step. In this way, the boundary of bistability beyond which a solution becomes unstable is found. See the Supplementary Information for more details.

\section*{Acknowledgment}

The author acknowledges funding from Alexander von Humboldt Foundation in the framework of the Sofja Kovalevskaja Award endowed by the German Federal Ministry of Education and Research.

\setcounter{figure}{0}
\setcounter{equation}{0}
\setcounter{section}{0}

\newcommand{\red}{\textcolor{red}}
\newcommand{\blue}{\textcolor{blue}}
\newcommand\figmainfirst{1}
\newcommand\figmainsecond{2}

\renewcommand\thesection{SI. \arabic{section}}
\renewcommand\thesubsection{\thesection.\arabic{subsection}}

\renewcommand\thefigure{SI.\arabic{figure}} 
\renewcommand\thetable{SI.\arabic{table}} 
\renewcommand\theequation{SI.\arabic{equation}} 
\makeatletter
\def\p@subsection{}
\makeatother
%\onecolumngrid
\clearpage

\begin{center}
	{\huge \bf 
		
		Supplementary Information for:}
	\\
	{\LARGE\bf	Evolution of cooperation and consistent personalities in public goods games\\ }
	\vspace{0.4cm}
	
	Mohammad Salahshour
	
\end{center}
\section{Methods}
\subsection{Overview of the model}
For completeness, here we give an overview of the model. We consider a population of $N$ individuals, who live for two rounds and possibly play a public goods game (PGG) in each round. Individuals have independent strategies in the two rounds. In the beginning of each generation (which we refer to as a time step), groups of size $g$ individuals are formed randomly from the population pool. In the first round, in each group, a PGG with enhancement factor $r_1$ is played. We consider two different scenarios. In the first scenario, those who cooperate in the first round, proceed to play a second PGG with all those who had cooperated in the first round in their group. Those who defect in the first round can not play a PGG for the second round. In the second scenario, both first round cooperators and first round defectors are able to play a PGG for the second round. However, in the second round, individuals are sorted based on their strategies in the first round. That is, all those who had cooperated in the first round in a group, form a subgroup to play a PGG (which we call the cooperative PGG), and all those who had defected in the first round form another subgroup to play their second PGG (which we call the defective PGG). We assume the second PGG to have an enhancement factor $r_2$.

Individuals gather payoff according to the outcome of the games. In addition, they receive a base payoff $\pi_0$ from other activities not related to the games. After the second round, a selection occurs, during which individuals are selected with a probability proportional to their payoff, such that the population size remains constant. That is each individual in the next generation is offspring to one of the individuals in the past generation with a probability proportional to its payoff. Offspring inherit the strategy of their parent. However, mutation in each strategy can occur with a probability $\nu$, in which case the value of the corresponding strategy is flipped to its opposite value.
\subsection{The simulations and numerical solutions}
Analytical solutions result from numerically solving the replicator dynamics of the models, derived in the next section. Simulations of the models are performed according to the model definition. Unless otherwise stated, both simulations and analytical solutions are performed with an initial condition in which all the strategies are found in similar frequencies in the population pool. For the solutions of the replicator dynamics, this is assured by setting the initial frequency of all the four strategies equal to $1/4$. For simulations, this is assured by a random assignment of the strategies in the beginning of the simulation. The phase diagram presented in Fig. (\ref{fig1rev}.a) is derived by locating the parameter values where a transition between different attractors occurs starting from a homogeneous initial condition. The boundary of bistabilities is derived by examining history dependence and checking for the existence of hysteresis in the evolution of the system. See section \ref{phasediagram} for more details.
\subsection{Replicator dynamics}
The model can be solved analytically in terms of replicator-mutation equations. These equations can be written as follows:
\begin{align}
\rho_{xy}(t+1)=\sum_{x',y'}\nu_{xy}^{x'y'}\rho_{x'y'}(t)\frac{\pi_{x'y'}(t)}{\sum_{x'',y''}\rho_{x''y''}(t)\pi_{x''y''}(t)}.
\label{eqrepM}
\end{align}
Here, $xy$ (as well as $x'y'$ and $x''y''$) refer to the strategies of the individuals, such that $x$ is the strategy of an individual in the first round, and $y$ is its strategy in the second round. $x$, $x'$ etc. can be either cooperation $C$ or defection $D$. $\nu_{xy}^{x'y'}$ is the mutation rate  from the strategy $x'y'$ to the strategy $xy$. These can be written in terms of mutation rate $\nu$ as follows: 
\begin{align}
\begin{cases}
\nu_{xy}^{x'y'}=1-2\nu+\nu^2 \quad &\textit{if\quad  ($x=x'$ \quad and \quad $y=y'$)},\\ 
\nu_{xy}^{x'y'}=\nu-\nu^2 \quad&\textit{if\quad ($x\neq x'$ \quad and \quad $y=y'$)\quad  or \quad ($x=x'$ \quad and \quad $y\neq y'$)}, \\
\nu_{xy}^{x'y'}=\nu^2\quad& \textit{if\quad ($x\neq x'$\quad and \quad$y \neq y'$)}.
\end{cases}
\end{align}
In eq. (\ref{eqrepM}), $\pi_{x'y'}$ is the expected payoff of an individuals with strategy $x'y'$. In the case of the first scenario, these are given by the following equations:
\begin{align}
\pi_{C_1C_2}=&\sum_{n_{D_1C_2}=0}^{g-1-n_{C_1C_2}-n_{C_1D_2}}\sum_{n_{C_1D_2}=0}^{g-1-n_{C_1C_2}}\sum_{n_{C_1C_2}=0}^{g-1}\bigg[r_1\frac{1+n_{C_1}}{g} +r_2\frac{1+n_{C_1C_2}}{1+n_{C_1}}\bigg]{\rho_{C_1C_2}}^{n_{C_1C_2}}{\rho_{C_1D_2}}^{n_{C_1D_2}}{\rho_{D_1C_2}}^{n_{D_1C_2}}\nonumber\\&{\rho_{D_1D_2}}^{g-1-n_{C_1C_2}-n_{C_1D_2}-n_{D_1C_2}}\binom{g-1}{n_{C_1C_2},n_{C_1D_2},n_{D_1C_2},g-1-n_{C_1C_2}-n_{C_1D_2}-n_{D_1C_2}}-2c+\pi_0,\nonumber\\
\pi_{C_1D_2}=&\sum_{n_{D_1C_2}=0}^{g-1-n_{C_1C_2}-n_{C_1D_2}}\sum_{n_{C_1D_2}=0}^{g-1-n_{C_1C_2}}\sum_{n_{C_1C_2}=0}^{g-1}\bigg[r_1\frac{1+n_{C_1}}{g} +r_2\frac{n_{C_1C_2}}{1+n_{C_1}}\bigg]{\rho_{C_1C_2}}^{n_{C_1C_2}}{\rho_{C_1D_2}}^{n_{C_1D_2}}{\rho_{D_1C_2}}^{n_{D_1C_2}}\nonumber\\&{\rho_{D_1D_2}}^{g-1-n_{C_1C_2}-n_{C_1D_2}-n_{D_1C_2}}\binom{g-1}{n_{C_1C_2},n_{C_1D_2},n_{D_1C_2},g-1-n_{C_1C_2}-n_{C_1D_2}-n_{D_1C_2}}-c+\pi_0,\nonumber\\
\pi_{D_1C_2}=&\sum_{n_{D_1C_2}=0}^{g-1-n_{C_1C_2}-n_{C_1D_2}}\sum_{n_{C_1D_2}=0}^{g-1-n_{C_1C_2}}\sum_{n_{C_1C_2}=0}^{g-1}\bigg[r_1\frac{n_{C_1}}{g} \bigg]{\rho_{C_1C_2}}^{n_{C_1C_2}}{\rho_{C_1D_2}}^{n_{C_1D_2}}{\rho_{D_1C_2}}^{n_{D_1C_2}}\nonumber\\&{\rho_{D_1D_2}}^{g-1-n_{C_1C_2}-n_{C_1D_2}-n_{D_1C_2}}\binom{g-1}{n_{C_1C_2},n_{C_1D_2},n_{D_1C_2},g-1-n_{C_1C_2}-n_{C_1D_2}-n_{D_1C_2}}+\pi_0,\nonumber\\
\pi_{D_1D_2}=&\sum_{n_{D_1C_2}=0}^{g-1-n_{C_1C_2}-n_{C_1D_2}}\sum_{n_{C_1D_2}=0}^{g-1-n_{C_1C_2}}\sum_{n_{C_1C_2}=0}^{g-1}\bigg[r_1\frac{n_{C_1}}{g} \bigg]{\rho_{C_1C_2}}^{n_{C_1C_2}}{\rho_{C_1D_2}}^{n_{C_1D_2}}{\rho_{D_1C_2}}^{n_{D_1C_2}}\nonumber\\&{\rho_{D_1D_2}}^{g-1-n_{C_1C_2}-n_{C_1D_2}-n_{D_1C_2}}\binom{g-1}{n_{C_1C_2},n_{C_1D_2},n_{D_1C_2},g-1-n_{C_1C_2}-n_{C_1D_2}-n_{D_1C_2}}+\pi_0.
\label{eqscenario1M}
\end{align}
Here, we have $n_{C_1}=n_{C_1C_2}+n_{C_1D_2}$. To write these equations, we used the fact that in a group with $n_{C_1C_2}$ individuals with strategy ${C_1C_2}$, and $n_{C_1D_2}$ individuals with strategy ${C_1D_2}$, $r_1\frac{1+n_{C_1}}{g}-c$ and $r_1\frac{n_{C_1}}{g}$ are, respectively, the expected payoff of an individual who cooperates, defects, in the first game. Those who defect in the first game do not gather payoff from the second game. On the other hand, those who cooperate in the first game, obtain a payoff from the second game as well (which can be negative or positive). This is $r_2\frac{1+n_{C_1C_2}}{1+n_{C_1}}-c$ for an individual with strategy $C_1C_2$, and $r_2\frac{n_{C_1C_2}}{1+n_{C_1}}$ for an individual with strategy $C_1D_2$. Finally, ${\rho_{C_1C_2}}^{n_{C_1C_2}}{\rho_{C_1D_2}}^{n_{C_1D_2}}{\rho_{D_1C_2}}^{n_{D_1C_2}}{\rho_{D_1D_2}}^{g-1-n_{C_1C_2}-n_{C_1D_2}-n_{D_1C_2}}$ $\binom{g-1}{n_{C_1C_2},n_{C_1D_2},n_{D_1D_2},g-1-n_{C_1C_2}-n_{C_1D_2}-n_{D_1C_2}}$, is the probability that a focal individual finds itself in a group with $n_{C_1C_2}$, $n_{C_1D_2}$, $n_{D_1C_2}$, and $n_{D_1D_2}$ individuals with, respectively, strategies $C_1C_2$, $C_1D_2$, $D_1C_2$, and $D_1D_2$. Here, $\binom{g-1}{n_{C_1C_2},n_{C_1D_2},n_{D_1D_2},g-1-n_{C_1C_2}-n_{C_1D_2}-n_{D_1C_2}}$ is the multinational coefficients (that is the number of ways that among the $g-1$ group mates of a focal individual, $n_{C_1C_2}$, $n_{C_1D_2}$, $n_{D_1C_2}$, $g-1-n_{C_1C_2}-n_{C_1D_2}-n_{D_1C_2}$ individuals have strategies, respectively, $C_1C_2$, $C_1D_2$, $D_1C_2$, and $D_1D_2$). Summation over all the possible configurations gives the expected payoff of the focal individual with the given strategy from the games. Finally, as all the individuals receive a base payoff $\pi_0$, this is added to the total payoff. Using the expressions in eq. (\ref{eqscenario1M}) for the expected payoff of different strategies in eq. (\ref{eqrepM}), we have a set of four equations which gives an analytical description of the model, in the limit of infinite population size. 
\begin{figure}%[ht]
	\centering
	\includegraphics[width=1\linewidth, trim = 80 221 65 5, clip,]{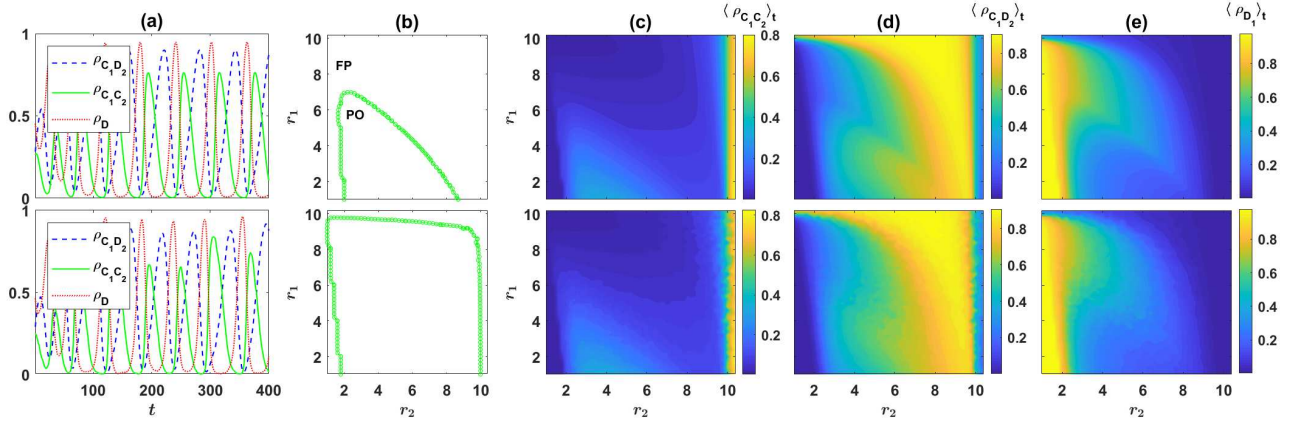}
	\caption{First Scenario Model (reward dilemma). (a) Density of different strategies as a function of time. Top panel represents numerical solutions of the replicator dynamics, and the bottom panel results from a simulation. The model shows cyclic dominance of different strategies. (b) The phase diagram of the model as resulted from solutions of the replicator dynamics, for two different mutation rates (top $\nu=10^{-3}$ and bottom $\nu=10^{-5}$). The dynamics settle in a periodic orbit or a fixed point depending on the parameters of the model. (c) to (e): Color plots of, respectively, $\langle\rho_{C_1C_2}\rangle_t$, $\langle\rho_{C_1D_2}\rangle_t$ and $\langle\rho_{D_1}\rangle_t=\langle\rho_{D_1C_1}+\rho_{D_1D_2}\rangle_t$. Top panels result from solutions of the replicator dynamics and bottom panels result from simulations. Cooperation evolves in the second game (c), and is maximized for moderate values of $r_2$. This renders entering the second game an incentive to cooperate in the first game, which promotes cooperation in the first game. The level of cooperation in the first game shows an overall increasing trend with increasing $r_2$. Here, $g=10$, $c=1$, and $\pi_0=2$. In (a), (c), (d), and (e) $\nu=10^{-3}$. In (a), $r_1=1.8$ and $r_2=3.8$. Simulation are performed in a population of size $N=5000$. For analytical solutions, the replicator dynamics is solved for $T=5000$ time steps and time averages are taken over the last $2000$ time steps. The simulation is performed for $T=3000$ time steps and the time averages are taken over the last $2500$ time steps.}
	\label{fig1}
\end{figure}

In the same way, it is possible to write down equations for the expected payoffs of individuals with different strategies in the second scenario. The difference with the preceding scenario is that, in the second scenario those who defect in the first round proceed to a second PGG as well. Thus, under the same notation and conventions as before, the individuals with strategies $D_1C_2$ and $D_1D_2$, obtain a payoff of, respectively, $r_2\frac{1+n_{D_1C_2}}{1+n_{D_1}}-c$ and $r_2\frac{n_{D_1C_2}}{1+n_{D_1}}$, from  their second game. Here, $n_{D_1}=n_{D_1C_1}+n_{D_1D_2}$. Thus, we have for the expected payoffs of different strategies in the second scenario:
\begin{align}
\pi_{C_1C_2}=&\sum_{n_{D_1C_2}=0}^{g-1-n_{C_1C_2}-n_{C_1D_2}}\sum_{n_{C_1D_2}=0}^{g-1-n_{C_1C_2}}\sum_{n_{C_1C_2}=0}^{g-1}\bigg[r_1\frac{1+n_{C_1}}{g} +r_2\frac{1+n_{C_1C_2}}{1+n_{C_1}}\bigg]{\rho_{C_1C_2}}^{n_{C_1C_2}}{\rho_{C_1D_2}}^{n_{C_1D_2}}{\rho_{D_1C_2}}^{n_{D_1C_2}}\nonumber\\&{\rho_{D_1D_2}}^{g-1-n_{C_1C_2}-n_{C_1D_2}-n_{D_1C_2}}\binom{g-1}{n_{C_1C_2},n_{C_1D_2},n_{D_1C_2},g-1-n_{C_1C_2}-n_{C_1D_2}-n_{D_1C_2}}-2c+\pi_0,\nonumber\\
\pi_{C_1D_2}=&\sum_{n_{D_1C_2}=0}^{g-1-n_{C_1C_2}-n_{C_1D_2}}\sum_{n_{C_1D_2}=0}^{g-1-n_{C_1C_2}}\sum_{n_{C_1C_2}=0}^{g-1}\bigg[r_1\frac{1+n_{C_1}}{g} +r_2\frac{n_{C_1C_2}}{1+n_{C_1}}\bigg]{\rho_{C_1C_2}}^{n_{C_1C_2}}{\rho_{C_1D_2}}^{n_{C_1D_2}}{\rho_{D_1C_2}}^{n_{D_1C_2}}\nonumber\\&{\rho_{D_1D_2}}^{g-1-n_{C_1C_2}-n_{C_1D_2}-n_{D_1C_2}}\binom{g-1}{n_{C_1C_2},n_{C_1D_2},n_{D_1C_2},g-1-n_{C_1C_2}-n_{C_1D_2}-n_{D_1C_2}}-c+\pi_0,\nonumber\\
\pi_{D_1C_2}=&\sum_{n_{D_1C_2}=0}^{g-1-n_{C_1C_2}-n_{C_1D_2}}\sum_{n_{C_1D_2}=0}^{g-1-n_{C_1C_2}}\sum_{n_{C_1C_2}=0}^{g-1}\bigg[r_1\frac{n_{C_1}}{g} +r_2\frac{1+n_{D_1C_2}}{1+n_{D_1}}\bigg]{\rho_{C_1C_2}}^{n_{C_1C_2}}{\rho_{C_1D_2}}^{n_{C_1D_2}}{\rho_{D_1C_2}}^{n_{D_1C_2}}\nonumber\\&{\rho_{D_1D_2}}^{g-1-n_{C_1C_2}-n_{C_1D_2}-n_{D_1C_2}}\binom{g-1}{n_{C_1C_2},n_{C_1D_2},n_{D_1C_2},g-1-n_{C_1C_2}-n_{C_1D_2}-n_{D_1C_2}}-c+\pi_0,\nonumber\\
\pi_{D_1D_2}=&\sum_{n_{D_1C_2}=0}^{g-1-n_{C_1C_2}-n_{C_1D_2}}\sum_{n_{C_1D_2}=0}^{g-1-n_{C_1C_2}}\sum_{n_{C_1C_2}=0}^{g-1}\bigg[r_1\frac{n_{C_1}}{g} +r_2\frac{n_{D_1C_2}}{1+n_{D_1}}\bigg]{\rho_{C_1C_2}}^{n_{C_1C_2}}{\rho_{C_1D_2}}^{n_{C_1D_2}}{\rho_{D_1C_2}}^{n_{D_1C_2}}\nonumber\\&{\rho_{D_1D_2}}^{g-1-n_{C_1C_2}-n_{C_1D_2}-n_{D_1C_2}}\binom{g-1}{n_{C_1C_2},n_{C_1D_2},n_{D_1C_2},g-1-n_{C_1C_2}-n_{C_1D_2}-n_{D_1C_2}}+\pi_0.
\end{align}
Using these expressions for the expected payoffs of individuals with different strategies in eq. (\ref{eqrepM}), we have the analytical description of the second scenario model, in the limit of infinite population size.
\begin{figure}[!ht]
	\centering
	\includegraphics[width=1\linewidth, trim = 142 15 120 28, clip,]{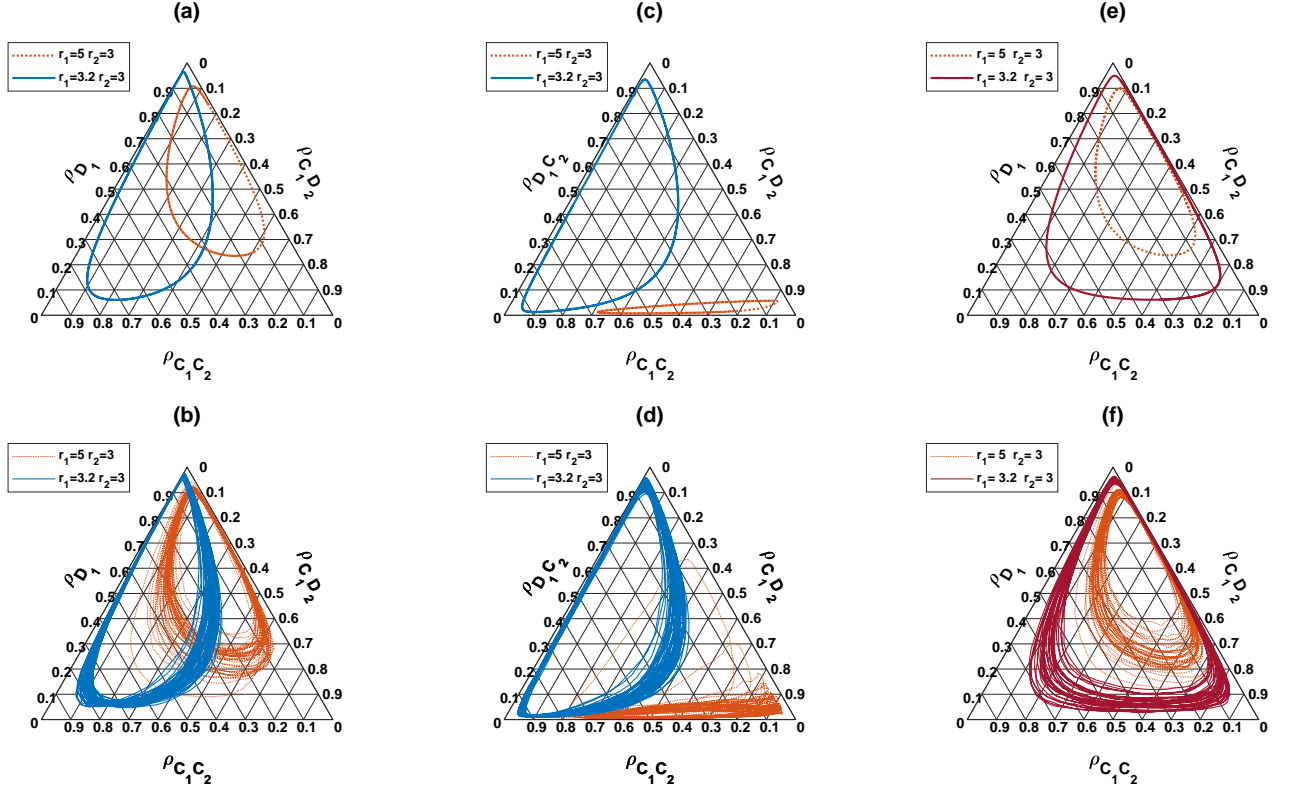}
	\caption{Periodic orbits. The periodic orbits in the second scenario model, as it results from the replicator dynamics (a and c), and from a simulation in a population of size $N=20000$ (b and d). (a) and (b), present the periodic orbits in the $\rho_{D_1}-\rho_{C_1D_2}-\rho_{C_1C_2}$ ternary diagram, and (c) and (d) present the same periodic orbit in $\rho_{D_1C_2}-\rho_{C_1D_2}-\rho_{C_1C_2}$ ternary diagram. The model shows two different periodic orbit, a defective periodic orbit (dotted red) for small $r_2$ and large $r_1$, and a cooperative periodic orbit (solid blue) for large $r_2$ and small $r_1$. (e) and (f) The periodic orbit in the first Model resulting from numerical solutions of the replicator dynamics (e), and a simulation in a population of size $N=20000$ (f). These correspond to the defective periodic orbit in the second scenario model. Here, $\nu=10^{-3}$, $g=10$, $c=1$, and $\pi_0=2$. In (b), (d), and (f), the simulations is run for $T=3000$ time steps and the last $2500$ time steps are shown.}
	\label{figSIcycle}
\end{figure}

\section{The first scenario: A reward dilemma solves the social dilemma}
Here, we study the first scenario, where only those who cooperate in the first round proceed to a second PGG. Depending on the parameters of the model, the dynamics can settle in either a fixed point or a periodic orbit (PO). We begin by studying the cyclic behavior. In Fig. (\ref{fig1}.a), the result of a numerical solution of the replicator dynamics (top), together with a simulation in a population of size $N=5000$ (bottom), are presented. Here, $g=10$, $\nu=10^{-3}$, $c=1$, $\pi_0=2$, $r_1=1.8$ and $r_2=3.8$. As can be seen, simulation in a finite population is in good agreement with the results of the replicator dynamics, which is an exact solution of the model for infinite population size.

We begin our analysis by noting that, in the cyclic phase, as can be seen in Fig. (\ref{fig1}), when the density of individuals who cooperate in the first round, and thus enter the second PGG is small, individuals can reach a high payoff by entering and cooperating in the second PGG. Thus $\rho_{C_1C_2}$ increases. When $\rho_{C_1C_2}$ increases enough, individuals can reach a higher payoff by defecting in the second PGG. At this point, $\rho_{C_1D_2}$ begins to increase, while $\rho_{C_1C_2}$ decreases. As the density of defectors in the second PGG increases, its profitability decreases, and thus individuals have no incentive to cooperate in the first round. Consequently, both $\rho_{D_1D_2}$ and $\rho_{D_1C_2}$, as well as $\rho_D=\rho_{D_1D_2}+\rho_{D_1C_2}$ increase, while other strategies decrease (we note that as those who defect in the first round do not enter the second game, the two strategies $D_1D_2$ and $D_1C_2$ are degenerate as they lead to the same payoff and are found in the same densities).

As mentioned before, cyclic behavior is not the only attractor of the system. This can be seen in Fig. (\ref{fig1}.b), where the phase diagram of the model for two different mutation rates (top $\nu=10^{-3}$ and bottom $\nu=10^{-5}$) is presented. For too large values of $r_1$, the dynamics settle in a fixed point (the region indicated by $FP$ in the figures). On the other hand, for smaller $r_1$, cyclic behavior occurs for intermediate values of $r_2$ (indicated by $PO$ in the figures). For both too large and too small $r_2$, a transition to a phase where the dynamics settle in a fixed point is observed. As can be seen by comparing the phase diagram for two different mutation rates, smaller mutation rates increase the size of the region where the dynamics settle in a periodic orbit.

\begin{figure}[!ht]
	\centering
	\includegraphics[width=1\linewidth, trim = 106 310 92 28, clip,]{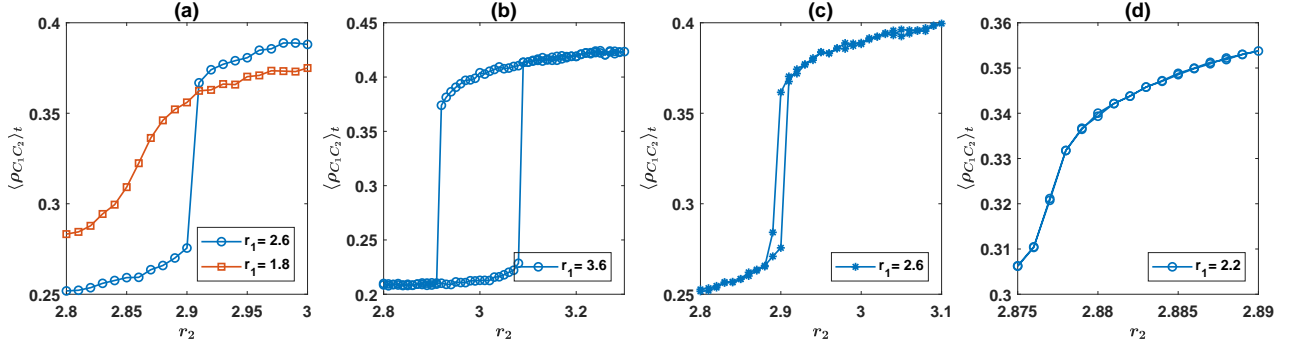}
	\caption{Bi-stability of the second scenario model. The time average of $\rho_{C_1C_2}$, as a function of $r_2$, for two fixed values of $r_1$ (as indicated in the figures), are plotted. While for small fixed $r_1$ the transition from the defective to the cooperative periodic orbit is gradual and shows no singularity, this transition becomes singular and happens discontinuously for large $r_1$. (b) to (d): The hysteresis loops for two different values of $r_1$. Here, the replicator equations are solved starting with a small value of $r_2$ below the transition. $r_2$ is slowly increased up to a large value above the transition and brought slowly back to its initial value. For large $r_1$ (b and d), depending on the path, the system equilibrates into different states. This shows the system is bistable close to the phase transition. However, for small $r_1$ (d), the system shows no bistability and the equilibrium state is history independent. Here, $\nu=10^{-3}$, $g=10$, $c=1$, and $\pi_0=2$, and the numerical solutions of the replicator equations are used. To derive the hysteresis loops, the replicator dynamics is solved starting with a homogeneous initial condition. After each $T$ time steps, the value of $r_2$ is changed by a small value (as shown in the figure). The values reported in the figure are averages in the last $\tau$ steps, where the system equilibrates. In (b) and (c), $T=5000$, and $\tau=3000$, and in (d), $T=20000$ and $\tau=5000$. In (a), the replicator dynamics is solved for $5000$ time steps, and time averages are taken over the last $2000$ steps.}
	\label{figSIHysteresis}
\end{figure}
To see how the cooperation level changes with $r_1$ and $r_2$, in Fig. (\ref{fig1}.c), Fig. (\ref{fig1}.d), and Fig. (\ref{fig1}.e), we plot, respectively, the time average of $\rho_{C_1C_2}$, $\rho_{C_1D_2}$ and $\rho_{D_1}=\rho_{D_1C_2}+\rho_{D_1D_2}$. For small $r_2$, such that the second PGG is not profitable enough to motivate cooperation in the first PGG, the dynamics settle into a defective fixed point, where the majority of the individuals defect in the first game and do not play the second game. Consequently $\rho_{D_1}$ takes a large value, and both $\rho_{C_1C_2}$ and $\rho_{C_1D_2}$ take small values. On the other hand, for very large $r_2$ the fixed point becomes a partially cooperative fixed point where most of the individuals cooperate in the first round, but defect in the second round. For $r_2$ larger than $g=10$, the second PGG is no longer a social dilemma and cooperation becomes the most rational strategy. This makes entering the second PGG a certain reward to cooperation in the first PGG, which promotes full cooperation in the first game.

An interesting question is how the level of cooperation in the two PGGs depends on the enhancement factors $r_1$ and $r_2$. As it is obvious in Fig. (\ref{fig1}), the level of cooperation in the first game shows an overall increasing trend with increasing $r_2$. This makes sense as with a higher value of $r_2$, entering the second game can provide a higher motivation for the individuals to cooperate in the first game. On the other hand, interestingly, the level of cooperation in the second game is maximized for a moderate value of $r_2$: increasing $r_2$ beyond this optimal value has a diverse effect on cooperation in the second game, as long as $r_2$ is small enough compared to the group size. That is, as long as we are in a region where the second round involves a social dilemma as well. This surprising behavior seems to result from the fact that by increasing $r_2$, a higher fraction of individuals cooperate in the first round to enter the second game. As the second PGG becomes more populated, maintaining cooperation in this game becomes more vulnerable to free riding and its level drops.
\section{The periodic orbits and transition between them}
\subsection{The periodic orbits in the first and second scenario model}
As mentioned in the main text, the second scenario can give rise to two qualitatively different periodic orbits, each stable in some range of the parameter values. Examples of these POs can be observed in the ternary plots in Fig. (\ref{figSIcycle}). Here, in each panel, two POs observed for two different values of $r_1$ and $r_2$, as specified in the figures are plotted. Here, $g=10$, $\nu=10^{-3}$, $c=1$, and $\pi_0=2$. In Fig. (\ref{figSIcycle}.a) and Fig. (\ref{figSIcycle}.b), the periodic orbits in the $\rho_{D_1}-\rho_{C_1D_2}-\rho_{C_1C_2}$ ternary diagrams are plotted (here, $\rho_{D_1}=\rho_{D_1C_2}+\rho_{D_1D_2}$). The same POs are plotted in the $\rho_{D_1C_2}-\rho_{C_1D_2}-\rho_{C_1C_2}$ ternary diagrams in Fig. (\ref{figSIcycle}.c) and Fig. (\ref{figSIcycle}.d). In Fig. (\ref{figSIcycle}.a) and Fig. (\ref{figSIcycle}.c), the analytical solutions are used, and in Fig. (\ref{figSIcycle}.b) and Fig. (\ref{figSIcycle}.d), a simulation in a population of size $N=20000$ is used.

For small $r_2$, in a fixed $r_1$ (or large $r_1$ with a fixed $r_2$) chosen below the the transition line, the defective PGG can not perform competitively compared to the cooperative PGG, as the vast majority of the individuals who defect in the first round and enter the defective PGG, defect in the second round as well. Consequently, $\rho_{D_1C_2}$ remains small. This PO is called the defective PO, and is plotted by red dotted line in Fig. (\ref{figSIcycle}.a) and Fig. (\ref{figSIcycle}.c). On the other hand, for larger $r_2$, in a fixed $r_1$ (or small $r_1$ in a fixed $r_2$) chosen above the transition, cooperation in both defective and cooperative PGGs evolves, and both $\rho_{D_1C_2}$ and $\rho_{C_1C_2}$ take large values. This PO is called the cooperative PO, and is plotted with solid blue line in Fig. (\ref{figSIcycle}.a) and Fig. (\ref{figSIcycle}.c). The results of a simulation in a population of size $N=20000$ for the same parameter values is presented in Fig. (\ref{figSIcycle}.b) and Fig. (\ref{figSIcycle}.d). Comparison reveals a good agreement between the results of a simulations in finite population size and the results of the replicator dynamics. However, in a finite population, population stochasticities can drive fluctuations in the periodic orbits.

In contrast, the first scenario only shows one type of PO, which coincides with the defective PO in the second scenario. This in turn, shows the competitive stability of a reward dilemma to promote cooperation in the face of a potential reward to defection. Two example POs resulting in the first scenario model are presented in Fig. (\ref{figSIcycle}.e). Here, $g=10$, $\nu=10^{-3}$, $c=1$, and $\pi_0=2$, and the replicator equations are used. To compare, in Fig. (\ref{figSIcycle}.f), two POs for the same parameter values, resulted from a simulation in a population of size $N=20000$ are plotted. Comparison reveals population stochasticities can result in fluctuations in the periodic orbits.

\begin{figure}%[ht]
	\centering
	\includegraphics[width=1\linewidth, trim = 76 140 40 64, clip,]{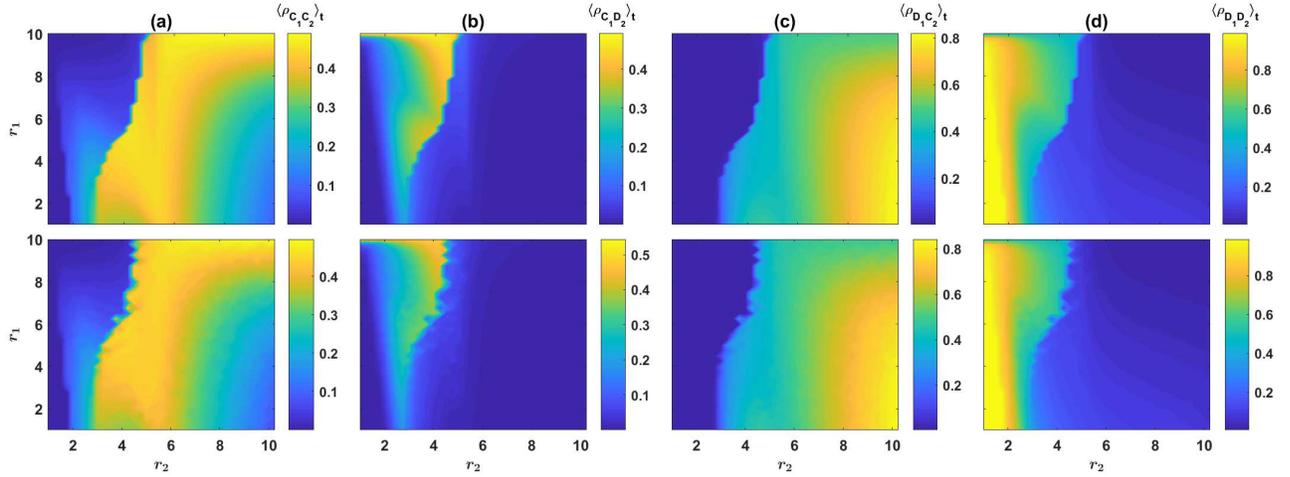}
	\caption{The density of different strategies in the second scenario. From (a) to (d), respectively, the time average of $\rho_{C_1C_2}$, $\rho_{C_1D_2}$, $\rho_{D_1C_2}$, and $\rho_{D_1D_2}$ are color plotted in the $r_1-r_2$ plane. Top panels result form numerical solution of the replicator dynamics, and bottom panels result from a simulation in a population of size $N=10000$. Here, $g=10$, $\nu=10^{-3}$, $c=1$, and $\pi_0=2$. For the analytical solutions, the replicator dynamics is solved for $T=5000$ time steps and the time average are taken over the last $2000$ time steps. The simulation is performed for $T=4000$ time steps and the time averages are taken over the last $3500$ time steps. }
	\label{figSI2}
\end{figure}
\subsection{Phase transition and cross-over between periodic orbits in the second scenario model}
We have seen that the second scenario model possesses two different types of periodic orbits, a defective periodic orbit for small $r_2$, and a cooperative periodic orbit for large $r_2$. This raises the question that whether a phase transition between these two periodic orbit exist? To answer this question, we need an order parameter, which takes different values for different attractors. We note that the time average of the density of any of the strategies in the population takes different values in the two different POs and can be used as an order parameter. Among these, we choose the density of individuals who cooperate in both games, $\rho_{C_1C_2}$. The time average of this quantity as a function of $r_2$, for two different values of $r_1$, is plotted in Fig. (\ref{figSIHysteresis}.a). Here, the replicator dynamics is solved using a homogeneous initial condition in which the density of all the strategies equals. As can be seen, for a fixed large $r_1$, as $r_2$ increases, the order parameter changes its value abruptly at a certain value of $r_2$. This shows the transition from the defective PO to the cooperative PO is discontinuous for large $r_1$. On the other hand, as can be seen in Fig. (\ref{figSIHysteresis}.a) (for $r_1=1.8$), for small $r_1$, as $r_2$ increases, there is a crossover from the defective PO to the cooperative PO without passing any singular transition.

The fact that the transition between the two periodic orbits is a discontinuous transition suggests the system can be bistable close to the transition. To see if this is the case, in Fig. (\ref{figSIHysteresis}.b) and Fig. (\ref{figSIHysteresis}.c), we drive the hysteresis loops, for two different values of $r_1$. Here, we have set $\nu=10^{-3}$, $g=10$, $c=1$, and $\pi_0=2$. To drive the hysteresis loops, in Fig. (\ref{figSIHysteresis}.b), we begin solving the replicator dynamics by setting $r_2=2.8$, and slowly increase $r_2$ up to $r_2=3.3$, and then decrease it back to the initial value of $r_2=2.8$. We plot the time average of $\rho_{C_1C_2}$ as a function of $r_2$ along the path. Here, for $r_2=2.8$, the dynamics settle into the defective PO, where $\langle\rho_{C_1C_2}\rangle_t$ takes a small value. As we gradually increase $r_2$ the system remains in the defective PO until reaching $r_2=3.08$. This can be attested by noting that in this interval the value of $\langle\rho_{C_1C_2}\rangle_t$ changes gradually. Beyond $r_2=3.08$, the defective PO becomes unstable and the system shows an abrupt transition into the cooperative PO. This can be seen in the abrupt change in the value of $\langle\rho_{C_1C_2}\rangle_t$. By increasing $r_2$ beyond this value, the system remains in the cooperative PO. However, when slowly decreasing $r_2$ from a large value to the initial value of $r_2=2.8$, the system follows a different path. On this path, the system remains in the cooperative PO even below the phase transition value of $r_2=3.08$. This shows the system is bistable for some range of $r_2$ below the phase transition value.

While the system shows hysteresis close to the onset of the cooperative PO for large $r_1$, the fact that for small $r_1$ there is a crossover between the two POs suggests the situation is different for small values of $r_1$. This can be clearly seen to be the case in Fig. (\ref{figSIHysteresis}.d), where $\langle\rho_{C_1C_2}\rangle_t$ is plotted using the same procedure. That is, the replicator dynamics is solved numerically starting with a small value of $r_2$. Slowly increasing $r_2$ up to a large value, $\langle\rho_{C_1C_2}\rangle_t$ slowly increases and the system shows a crossover from the defective PO to the cooperative PO. Decreasing $r_2$ slowly from a large value of $r_2$ back to its initial value, the system follows the same path. This shows no bistability and hysteresis is at work for small values of $r_1$.

\subsection{Derivation of the phase diagram and boundaries of bistability}
\label{phasediagram}
To drive the phase diagram of the second scenario model, we solve the replicator dynamics using a homogeneous initial condition (i.e. $\rho_x=0.25$, for $x={C_1C_2}$, ${C_1D_2}$, $D_1C_2$, and ${D_1D_2}$), and for each fixed $r_1$ identify the value of $r_2$ where the system shows a phase transition from a fixed point to the defective periodic orbit (blue line in Fig. (\ref{fig1rev}.a) in the main text). Similarly, we identify the value of $r_2$ where a transition to the cooperative periodic orbit occurs (red line in Fig. (\ref{fig1rev}.a) in the main text). The phase transition line between the defective and cooperative POs in Fig. (\ref{fig1rev}.a) is defined as the region where, starting from a homogeneous initial condition, a transition between these two equilibrium states occur.

To derive the boundaries of bistability in Fig. (\ref{fig1rev}.a), by solving the replicator dynamics for different initial conditions, we check that the transition from the fixed point to the defective periodic orbit shows no bistability and does not depend on the initial condition. Thus, this line coincides with the blue line in Fig. (\ref{fig1rev}.a). Similarly, the transition form the cooperative periodic orbit to the fixed point for large $r_2$ shows no bistability and occurs in the same parameter values for all the initial conditions. On the other hand, for large enough $r_1$, the onset of the cooperative periodic orbit when increasing $r_2$ shows bistability. To derive the lower boundary of the bistability region (that is the value of $r_2$ below which the cooperative PO is unstable), we solve the replicator dynamics starting with a large value of $r_2$, above the defective-cooperative PO transition, such that the dynamics settle into the cooperative periodic orbit. We then slowly decrease $r_2$ in small steps. In each step, as the initial condition, we use the equilibrium state of the dynamics in the last step (i.e. the attractor of the system for a slightly larger $r_2$). In this way, we are able to find the lower boundary of the bistable region as the largest value of $r_2$ below which the cooperative PO becomes unstable. We note that this procedure is necessary for deriving the lower boundary of the bistability region, as below the phase transition, although stable, the cooperative PO has a very small basin of attraction. Thus, using different randomly chosen initial conditions, this periodic orbit can remain unobserved.

To derive the upper boundary above which the defective periodic orbit, or the fixed point becomes unstable, similarly to the previous procedure, we start by numerically solving the replicator dynamics for a small enough value of $r_2$. Then we increase the value of $r_2$ ‌in small steps, using the equilibrium state of the dynamics for a slightly smaller $r_2$ as the initial condition for the next step. This allows us to identify the smallest value of $r_2$ above which the defective periodic orbit or the fixed point becomes unstable as the upper boundary of the bistable region. Interestingly, for small enough $r_1$, The upper boundary of the bistable region coincides with the phase transition line.

To derive the phase diagram in the first scenario model, presented in Fig. (\ref{fig1}.b), we numerically solve the replicator dynamics to determine different equilibrium states of the system. By solving the replicator dynamics for different initial conditions, we check that the system shows no bistability, and thus, the equilibrium state does not depend on the initial conditions.

\begin{figure}%[ht]
	\centering
	\includegraphics[width=1\linewidth, trim = 118 290 70 20, clip,]{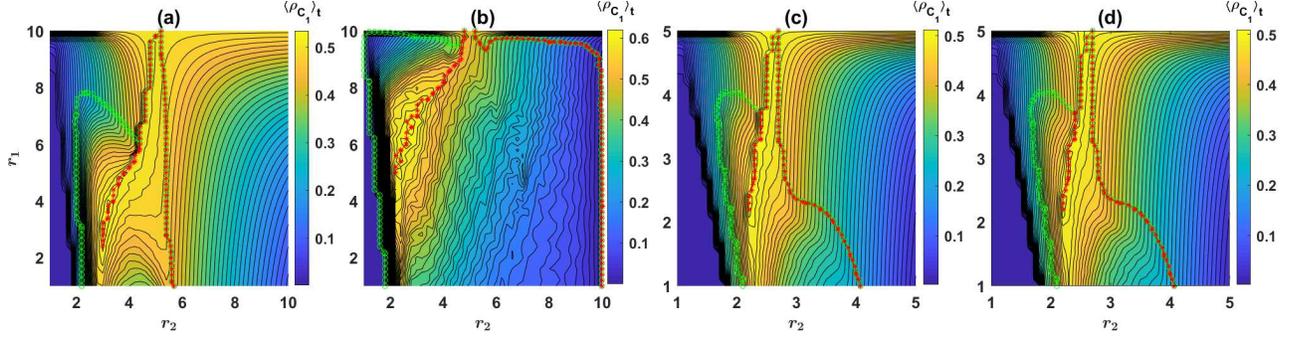}
	\caption{Cooperation level in the first round in the second scenario model. Contour plot of $\langle\rho_{C_1}\rangle_t=\langle\rho_{C_1C_2}+\rho_{C_1D_2}\rangle_t$ over $r_1-r_2$ plane. The lines of phase transitions are marked with green and red markers. For each fixed $r_1$, $\langle\rho_{C_1}\rangle_t$ is maximized for an intermediate value of $r_2$, which coincides with the edge of bistability. In (a) $g=10$, $\nu=10^{-3}$, in (b) $g=10$, $\nu=10^{-5}$, and in (c) and (d) $g=5$, $\nu=10^{-3}$. In all the cases the replicator dynamics is solved for $T=5000$ time steps, and time averages are taken over the last $2000$ time steps. In (a) to (c) the initial condition is a homogeneous initial condition in which the initial density of all the strategies equals, and in (d), the $r_1-r_2$ plane is divided into small cells of linear size $0.1$, and a randomly chosen initial condition is used for each cell (as described in the text). In all the cases $c=1$ and $\pi_0=2$.}
	\label{figSIoptimal}
\end{figure}
\section{The density of different strategies in the second scenario model}
The density of different strategies in the second scenario are plotted in Fig. (\ref{figSI2}). The time average density of strategies, $\rho_{C_1C_2}$, $\rho_{C_1D_2}$, $\rho_{D_1C_2}$, $\rho_{D_1D_2}$ are plotted respectively, in Fig. (\ref{figSI2}.a), Fig. (\ref{figSI2}.b), Fig. (\ref{figSI2}.c), and Fig. (\ref{figSI2}.d). Top panels result from numerical solutions of the replicator dynamics, and bottom panels result from a simulation in a population of size $N=10000$. Here, $g=10$, $\nu=10^{-3}$, $c=1$, and $\pi_0=2$. As can be seen, the results of replicator dynamics, an exact solution of the model in the infinite population limit, are in good agreement with simulations in finite population size.

For too small values of $r_2$, the system settles into a defective fixed point with small fraction of cooperators. As $r_2$ increases, at a certain value of $r_2$ the cooperative PGG starts to attract individuals. Consequently, the density of the strategies $C_1C_2$ and $C_1D_2$ start to increase. This is the region of the phase diagram where the defective PO is stable. In this region, by increasing $r_1$, $\rho_{C_1C_2}$ decreases, while $\rho_{C_1D_2}$ increases. This shows that while higher enhancement factor of the first game may have a positive effect on the cooperation level in this game, it can have an adverse effect on cooperation in the second game.

By further increasing $r_2$, the defective PGG starts to attract individuals as well. That is, those who defect in the first round, start to cooperate in the second round as well, and consequently, the defective PGG starts to yield positive reward. At this point, the defective PO becomes unstable, and the dynamics settle into the cooperative PO. In this region, the density of individuals who defect in the second round significantly decreases, and the vast majority of the individuals cooperate in the second round. This is due to the fact that competition between the cooperative and defective PGGs promotes cooperation in the second round. By increasing $r_2$ in this region, $\rho_{D_1C_2}$ increases while the density of the other strategies, including $\rho_{C_1C_2}$ decrease. Thus, when $r_2$ is too large, the cooperation level in the first game decreases. This in turn results from the same reason that promotes cooperation in the second round: As the competition between cooperative and defective PGGs promote cooperation in the second round, individuals receive a high payoff from the second round irrespective of what they do in the first round and which PGG they enter for the second round. In such circumstances, defection in the first round, which is the rational choice prevails.

Overall, this analysis shows enhancement factors of the game in the two rounds, can have complicated and surprising effect on the cooperation level in these two games. A point to which we will return in the next section.

\begin{figure}%[ht]
	\centering
	\includegraphics[width=1\linewidth, trim = 118 290 40 20, clip,]{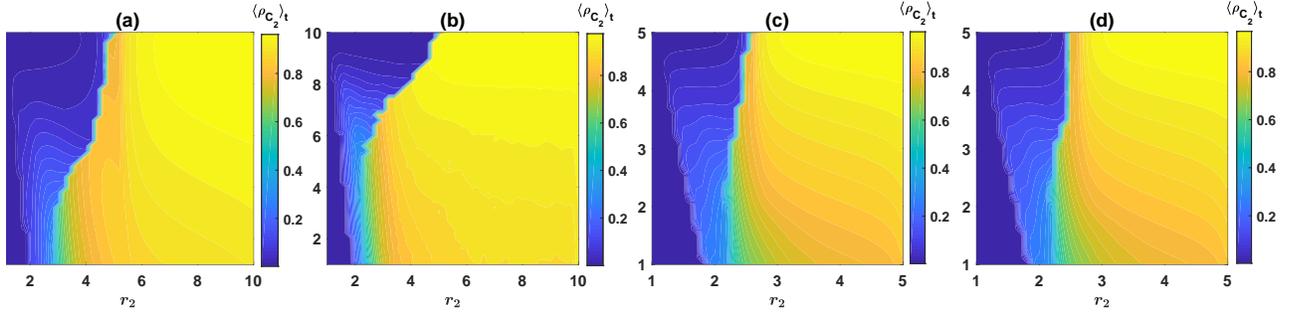}
	\caption{Cooperation level in the second round in the second scenario model. Contour plot of $\langle\rho_{C_2}\rangle_t=\langle\rho_{C_1C_2}+\rho_{D_1C_2}\rangle_t$ over $r_1-r_2$ plane. The cooperation level in the second game significantly increases above the transition to the cooperative PO, and monotonically increases by increasing $r_2$ above this transition. In (a), $g=10$, $\nu=10^{-3}$, in (b), $g=10$, $\nu=10^{-5}$, and in (c) and (d), $g=5$, $\nu=10^{-3}$. In all the cases the replicator dynamics is solved for $T=5000$ time steps, and time averages are taken over the last $2000$ time steps. In (a) to (c) the initial condition is a homogeneous initial condition in which the initial density of all the strategies equals, and in (d), the $r_1-r_2$ plane is divided into small cells of linear size $0.1$, and a randomly chosen initial condition is used for each cell (as described in the text). In all the cases $c=1$ and $\pi_0=2$.}
	\label{figSIcop2}
\end{figure}
\section{Dependence of the cooperation level on parameters}
In Fig. (\ref{figSIoptimal}.b) to Fig. (\ref{figSIoptimal}.d), we plot the contour plot of the time average cooperation level in the first game, $\langle\rho_{C_1}\rangle_t=\langle\rho_{C_1C_2}+\rho_{C_1D_2}\rangle_t$, in the $r_1-r_2$ plane for different parameter values. In Fig. (\ref{figSIoptimal}.a), $g=10$ and $\nu=10^{-3}$, in Fig. (\ref{figSIoptimal}.b), $g=10$ and $\nu=10^{-5}$, and in Fig. (\ref{figSIoptimal}.c), $g=5$ and $\nu=10^{-3}$. In all the cases $\pi_0=2$. In these figures, we have also indicated the position of the phase transitions. As can be seen, the same phases appear for different mutation rates and group sizes. However, parameter values can have quantitative effects on the position of the phase transitions and the cooperation evel. Particularly, smaller mutation rates increase the region of the phase space where the dynamics show cyclic behavior.

Interestingly, for a fixed $r_1$, cooperation level in the first game is maximized for an intermediate value of $r_2$, such that increasing $r_2$ beyond this value has a detrimental effect on the level of cooperation in the first game. This is due to the fact that for larger values of $r_2$ the cooperation level in both cooperative and defective PGGs increase. This decreases the difference in the quality of these two PGGs. While for smaller values of $r_2$, cooperation level in the defective PGG is much smaller than that in the cooperative PGG. This motivates individuals to cooperate in the first round in order to enter the cooperative PGG. The maximum cooperation level in the first game is achieved exactly on the transition line between the two periodic orbits, which as shown before, coincides with the edge of bistability. 

To see how the initial conditions affect the cooperation level, in Fig. (\ref{figSIoptimal}.d), we solve the replicator dynamics using different randomly chosen initial conditions. Here, the $r_1-r_2$ plane is divided into small cells of linear size $0.1$, and for each cell a randomly chosen initial condition is used to solve the replicator dynamics. For the initial conditions, we set the initial densities equal to $\rho_{C_1C_2}=a_{C_1C_2}/d$, $\rho_{C_1D_2}=a_{C_1D_2}/d$, $\rho_{D_1C_2}=a_{D_1C_2}/d$, and $\rho_{D_1D_2}=a_{D_1D_2}/d$, where $d=a_{C_1C_2}+a_{C_1D_2}+a_{D_1C_1}+a_{D_1D_2}$, and $a_{C_1C_2}$, $a_{C_1D_2}$, $a_{D_1C_2}$, and $a_{D_1C_2}$, are random numbers drawn uniformly at random in the interval $[0,1]$. Here, the phase diagram (which is derived using a homogeneous initial condition) is superimposed as well. As can be seen, the initial conditions does not affect the equilibrium state of the system: For all the randomly chosen initial conditions, the dynamics settle in the same equilibrium state as for a homogeneous initial condition. Surprisingly, this is also valid in the bistable region. This shows that, although in the bistable region, different equilibrium states are possible, non-equilibrium states have a very small basin of attraction, such that small deformations in a non-equilibrium state can cause the system to settle into the equilibrium state.

Finally, we study the dependence of the cooperation level in the second game on the parameters. For this purpose, in Fig. (\ref{figSIcop2}.a) to Fig. (\ref{figSIcop2}.c), we plot the time average cooperation level in the second game, $\langle\rho_{C_2}\rangle_t=\langle \rho_{C_1C_2}+\rho_{D_1C_2}\rangle_t$ in the $r_1-r_2$ plane. In Fig. (\ref{figSIcop2}.a), $g=10$ and $\nu=10^{-3}$, in Fig. (\ref{figSIcop2}.b), $g=10$ and $\nu=10^{-5}$, and in Fig. (\ref{figSIcop2}.c), $g=5$ and $\nu=10^{-3}$. In all the cases, for small values of $r_2$, the dynamics settle into, either a fixed point, or the defective periodic orbit. In both cases, cooperation does not evolve in the defective PGG. This keeps the cooperation level small in this region. However, in the region of the phase diagram where the dynamics settle into the cooperative PO, where cooperation in both the defective PGG and the cooperative PGG evolves, cooperative level in the second game significantly increases. In this phase, the cooperation level in the second game monotonically increases by increasing $r_2$. It is important to notice the situation was completely different in the second scenario model, where only first round cooperators proceed to play a second PGG. As we have seen, in that case, cooperation level in the second game is optimized for an intermediate value of $r_2$, and increasing $r_2$‌beyond this optimal value decreases cooperation level in the second game. The different situation in the second scenario model comes from the fact that when first round defectors are allowed to play a second PGG as well, competition between the social (cooperative) and anti-social (defective) PGGs increases, and maintains cooperation in the second round, even when the temptation to free ride is high. This result shows, a potential reward to defection, surprisingly, can further stabilize cooperation.
\begin{figure}%[ht] 
	\centering
	\includegraphics[width=1\linewidth, trim = 20 135 11 60, clip,]{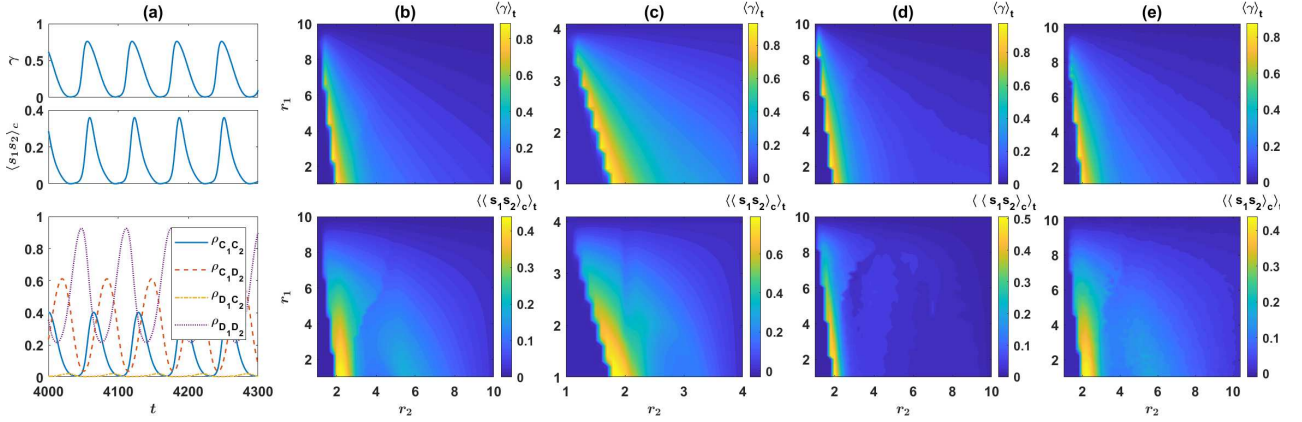}
	\caption{Evolution of consistent personalities. (a) The measure of consistency in personality, defined as $\gamma=[P(C_2|C_1)+P(D_2|D_1)-P(D_2|C_1)-P(C_2|D_1)]/2$ (top), and the connected correlation function between strategies of individuals in the two rounds, $\langle s_1s_2\rangle_c=\langle s_1s_2\rangle-\langle s_1\rangle\langle s_2\rangle$ (middle), as a function of time in the stationary state. The density of different strategies is plotted as well (down). Both measures remain always non-negative. This shows individuals are more likely to have the same strategies in the two rounds, and consistent cooperative or defective personalities evolve. (b), (c), (d) and (e): Contour plot of the time average of $\gamma$ (top), and the time average of $\langle s_1s_2\rangle_c$ (bottom), in the $r_1-r_2$ plane, for different parameter values. In (a) to (d), the replicator dynamics are numerically solved for $T=5000$ time steps, and time averages are taken over the last $2000$ time steps. In (e), a simulation on a population of size $N=5000$ for $T=3000$ time steps is performed, and the time averages are taken after discarding the first $500$ time steps. In (a), (b), and (e) $g=10$ and $\nu=10^{-3}$, in (c) $g=4$ and $\nu=10^{-3}$, and in (d) $g=10$ and $\nu=10^{-5}$. In (a) $r_1=2.8$ and $r_2=4.8$. In all the cases $c=1$ and $\pi_0=2$.}
	\label{figSIpersonality}
\end{figure}
\section{The evolution of cooperative and defective personalities}
As argued in the main text, in our model with the second scenario where both first round cooperators and first round defectors need to choose a strategy in the second round, individuals are more likely to have similar strategies in the two rounds than it can occur by chance. This shows individuals develop consistent personalities in the course of evolution. Here, we show that this result is robust for different parameter regimes. 

To see that individuals tend to evolve consistent personalities, we consider two different measures of personality consistency. The first measure is based on the conditional probability that an individual has strategy $s_2$ in the second round, given that it has strategy $s_1$ in the first round, $P(s_2|s_1)$. As the first measure of the consistency of individual's strategies in the two rounds, we define the personality consistency measure, $\gamma$, as $\gamma=[P(C_2|C_1)+P(D_2|D_1)-P(D_2|C_1)-P(C_2|D_1)]/2$. As a second measure, we consider the connected correlation function between the strategies of the individuals in the two rounds $\langle s_1s_2\rangle_c=\langle s_1s_2\rangle-\langle s_1\rangle\langle s_2\rangle$. Here, $\langle . \rangle$ denotes an average over the population. To calculate this, we assign $-1$ to the strategy $D$, and $+1$ to the strategy $C$.

Both measures always lie between $-1$ and $+1$. A positive value shows that individuals are more likely to have consistent strategies in the two rounds, and a negative value indicates that individuals are more likely to have opposite strategies in the two rounds. In Fig. (\ref{figSIpersonality}.a), we plot $\gamma$ (top panel) and $\langle s_1s_2\rangle_c$ (middle panel) in the stationary state, as a function of time. Here, the replicator dynamics is used, and $g=10$, $\nu=10^{-3}$, $c=1$, $\pi_0=2$, $r_1=2.8$ and $r_2=4.8$. As can be seen, both $\gamma$ and $\langle s_1s_2\rangle_c$ follow a cyclic behavior. Importantly, When the density of consistent $C_1C_2$ and $D_1D_2$ strategies are high, $\gamma$ and $\langle s_1s_2\rangle_c$ take a large value and they drop as inconsistent $C_1D_2$ and $D_1C_2$ strategies reach a high fraction. Furthermore, $\gamma$ and $\langle s_1s_2\rangle_c$ always remain non-negative. This shows strategies always remain consistent. To see how the consistency of the strategies change in the whole parameter regime, in Fig. (\ref{figSIpersonality}.b) to Fig. (\ref{figSIpersonality}.d), we plot the time average of $\gamma$ (top panels) and $\langle s_1s_2\rangle_c$ (bottom panels), over $r_1-r_2$ plane. In Fig. (\ref{figSIpersonality}.a), $g=10$ and $\nu=10^{-3}$, in Fig. (\ref{figSIpersonality}.c) $g=4$ and $\nu=10^{-3}$, and in Fig. (\ref{figSIpersonality}.d), $g=10$ and $\nu=10^{-5}$. In all the case $c=1$ and $\pi_0=2$. Here, the replicator-mutation equations are solved for $T=5000$ time steps, and time averages are taken over the last $2000$ time steps. As can be seen, in all the cases both measures remain non-negative. For small $r_2$, such that the system remains in the defective fixed point and $D_1D_2$ strategies prevail, both measures take a small value close to zero. This shows that in this regime, although strategies seem to be consistent, however, there is no correlation between the strategies in the two rounds, and thus our measures predict no personality consistency. As the frequency of non-defective strategies start to increase, both measures start to become large, which shows the existent strategies start to become consistent.

To see that the results agree with simulations in finite populations, in Fig. (\ref{figSIpersonality}.e) a simulation on a population of size $N=5000$ is used. Here $g=10$, $\nu=10^{-3}$, $c=1$ and $\pi_0=2$. This corresponds to the parameter values used in Fig. (\ref{figSIpersonality}.b). As can be seen, our results are valid in finite populations as well. In addition, a good agreement can be seen between analytical predictions, valid in the infinite population limit, and the simulation results in finite population size.

\end{document}